\documentclass{pj}
\usepackage{upgreek}
\usepackage{combelow}
\usepackage{comment}
\usepackage{cite}
\usepackage{tikz,graphicx}
\usepackage{amsmath}
\usepackage{color}
\usepackage[mathscr]{eucal}
\usepackage{amssymb}

\def\XXint#1#2#3{{\setbox0=\hbox{$#1{#2#3}{\int}$}
     \vcenter{\hbox{$#2#3$}}\kern-.5\wd0}}

\newcommand{\be}[1]{\begin{equation} \label{#1} }
\newcommand{\bea}[1]{\begin{eqnarray} \label{#1} }

\newcommand{\bfi}{\begin{figure}}
\newcommand{\efi}{\end{figure}} 
\newcommand{\ee}{\end{equation}}
\newcommand{\eea}{\end{eqnarray}}
\newcommand{\bib}{\bibitem}
\newcommand{\lbl}{\label}

\newcommand{\w}{{\omega}}

\newcommand{\vk}{{\bf k}}

\newcommand{\va}{{\bf a}}
\newcommand{\vb}{{\bf b}}
\newcommand{\vc}{{\bf c}}

\newcommand{\vv}{{\bf v}}
\newcommand{\vf}{{\bf f}}
\newcommand{\vp}{{\bf p}}
\newcommand{\vm}{{\bf m}}

\newcommand{\vn}{{\bf n}}
\newcommand{\vzh}{{\bf \hat{z}}}
\newcommand{\vyh}{{\bf \hat{y}}}
\newcommand{\vxh}{{\bf \hat{x}}}
\newcommand{\vrh}{{\bf \hat{r}}}

\newcommand{\vRh}{{\bf \hat{R}}}
\newcommand{\vr}{{\bf r}}
\newcommand{\vH}{{\bf H}}
\newcommand{\vE}{{\bf E}}

\newcommand{\vA}{{\bf A}}

\newcommand{\vF}{{\bf F}}

\newcommand{\vJ}{{\bf J}}
\newcommand{\vG}{{\bf G}}

\newcommand{\vR}{{\bf R}}
\newcommand{\vC}{{\bf C}}

\newcommand{\dT}{\overline{\bf T}}
\newcommand{\dI}{\overline{\bf I}}
\newcommand{\dQ}{\overline{\bf Q}}

\newcommand{\vnh}{{\bf \hat{n}}}

\newcommand{\vrho}{{\mbox{\boldmath $\rho$}}}
\newcommand{\vphih}{{\bf \hat{\mbox{\boldmath {$\phi$}}}}}
\newcommand{\vthh}{{\bf \hat{\mbox{\boldmath {$\theta$}}}}}

\newcommand{\vOmega}{\overline{\mbox{\boldmath $\Omega$}}}
\newcommand{\la}{\lesssim}

\newcommand{\eps}{\epsilon}

\begin{document}
\setcounter{page}{1}
%\pjheader{Vol.\ x, y--z, 2018}

\title[]%Force and Hidden Momentum] 
{\mbox{}\\[-20mm]Force and Hidden Momentum for Classical Microscopic Dipoles}
%\footnote{\it Received date}  \footnote{\hskip-0.12in*\, Corresponding
%author:~Arthur~D.~Yaghjian (a.yaghjian@comcast.net).}
%\footnote{\hskip-0.12in\textsuperscript{} The author works as an Electromagnetics Research Consultant,
%115 Wright Road, Concord, MA 01742, USA.}
%
\vspace{-2mm}
\author{Arthur~D.~Yaghjian}%\textsuperscript{*}}
\runningauthor{}%Yaghjian}
%\begin{document}
\footnote{}  \footnote{}
\footnote{}
%\setcounter{page}{1}
%\pjheader{Vol.\ x, y--z, 2018}
%\pjheader{}
%\title{Force and Hidden Momentum for Classical Microscopic Dipoles}
%
%\author{Arthur~D.~Yaghjian\textsuperscript{*}}
%
%\runningauthor{Yaghjian}
%
%\tocauthor{Arthur~Yaghjian}
%\setcounter{page}{1} 
%\pjheader{Vol.\ x, y--z, 2018}
%\jheader{Vol.\ x, y--z, 2013}
%\pheader{Vol.\ x, y--z, 2013}
%
%
\vspace{-2mm}
%{\Large\bf Force and Hidden Momentum for Classical Microscopic Dipoles}\\[8mm] \centerline{\bf Arthur D. Yaghjian}
\par
\begin{abstract}
The concept of hidden momentum is reviewed and the first rigorous derivation from Maxwell's equations is provided for the electromagnetic force on electrically small perfect electric conductors of arbitrary shape in {\color{black} bandlimited but otherwise} arbitrarily time-varying fields.  It is proven for the Amperian magnetic dipoles of these perfect conductors that a ``hidden-momentum" electromagnetic force exists that makes the force on these time varying Amperian magnetic dipoles equal to the force on magnetic-charge magnetic dipoles with the same time varying magnetic dipole moment in the same time varying externally applied fields.  The exact Mie solution to the perfectly conducting sphere under plane-wave illumination is used to prove that the expressions for the total and hidden-momentum forces on the arbitrarily shaped electrically small perfect conductors correctly predict the forces on perfectly conducting spheres.  {\color{black} Remarkably, it is found that the quadrupolar fields at the surface of the sphere are required to obtain the correct total force on the sphere even though the quadrupolar moments are negligible compared to the dipole moments as the electrical size of the sphere approaches zero.} %Lastly, Newton's third law of motion for sources in each other's quasistatic fields is used to argue that all passive, lossy or lossless, microscopic Amperian dipoles exhibit the hidden-momentum force that makes the total force on Amperian and magnetic-charge magnetic dipoles with the same magnetic dipole moment equal in the same externally applied fields.  The Mie solution for the force on electrically small lossy spheres is used to further confirm these general results for the hidden momentum argued from Newton's third law.
\end{abstract}
%
%\mytableofcontents
%\tableofcontents
%
\section{Introduction}
\begin{figure}[ht]
\mbox{}\\[-9mm]
\begin{center}
\includegraphics[width =6.0in]{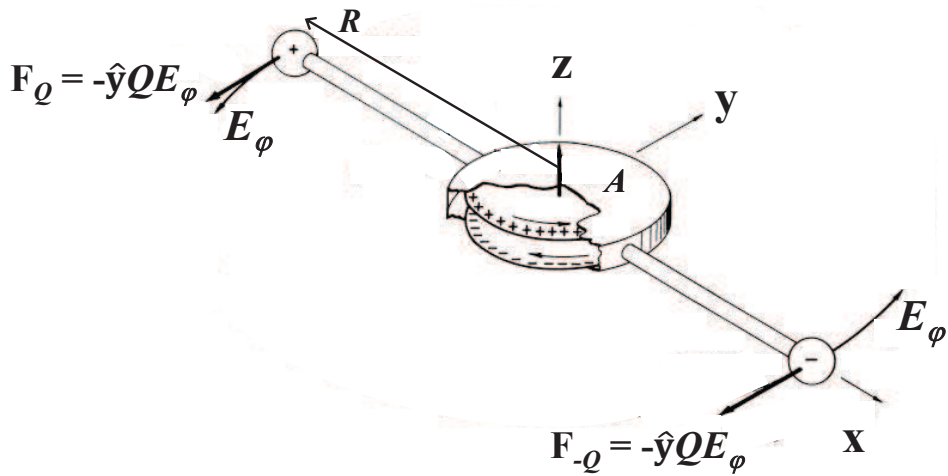}
\end{center}
\mbox{}\\[-14mm]
\caption{\label{Shockley}Counter-rotating disks, with oppositely charged rims, inside a pillbox having equal length radial arms with equal and opposite point charges ($\pm Q$) at their ends.  In the idealized limit the space between the two disks approaches zero so that the net charge density of the two disks approaches zero.}
\end{figure}

\lbl{Intro}
In their 1967 paper \cite{S&J}, Shockley and James considered two equal but opposite point charges ($\pm Q$) connected by radial arms extending to a radius $R$ on opposite sides of a circular pillbox containing two counter-rotating disks of area $A$ with oppositely charged rims so as to create circulating current and an Amperian magnetic dipole moment; see Fig. \ref{Shockley}.  The disks fit snugly but without friction against the pillbox.   
\mbox{}\\[3mm]
Assuming a vanishingly small frictional force between the counter-rotating disks that slowly (to avoid radiated fields) brings the disks to rest with a current $I(t)$ that slowly approaches zero, Shockley and James begin by determining the impulse (electromagnetic force integrated over time) imparted to the two opposite point charges by the magnetic dipole moment $\vm(t)= I(t)A\vzh$ as it decays to zero.  Specifically, the magnetic dipole moment produces an electric field $E_\phi$ on the two point charges that is dominated by the inductive electric field, $E_\phi(t) = -(\mu_0 A/4\pi R^2) dI(t)/dt$. This time-domain electric field, which can be obtained by taking the Fourier transform of the corresponding frequency-domain electric field of the magnetic dipole \cite[eq. (37), p. 437]{Stratton}, exerts a force equal to $-QE_\phi(t)\vyh$ on each point charge so that the total force on the two point charges is $\vF_{\pm Q}(t) =\vyh(Q\mu_0 A/2\pi R^2) dI(t)/dt$.  Integrating over time from $t=0$ (when the disks begin to slow down) to $t=\infty$ (when the disks are stationary), gives the electromagnetic impulse imparted to the two point charges, namely
\be{S1}
\vG_Q = \int\limits_0^\infty\vF_{\pm Q}(t)dt = -\vyh\frac{Q\mu_0 I(0)A}{2\pi R^2} = -\frac{1}{c^2}\vm(0)\times\vE_0
\ee
where $\vE_0$ is the Coulomb electric field at the center of the pillbox produced by the two point charges  and the speed of light is $c=1/\sqrt{\mu_0\eps_0}$ with $\mu_0$ and $\eps_0$ equal to the permeability and permittivity of free space.
In deriving (\ref{S1}), the magnetic dipole is assumed concentrated at the center of the pillbox and the time delay over the distance $R$ from the center of the pillbox to the two point charges is assumed negligible for the slowly decaying magnetic moment, that is, $E_\phi(t-R/c)\approx E_\phi(t)$.  Since the counter-rotating disks contain no net charge density as their spacing approaches zero, the electromagnetic force exerted on the disks by the electric field of the two point charges is zero.\footnote{As the disks slow down, there is also a force on $\vJ(\vr,t)$ from its own magnetic field $\vH(\vr,t)$ given by $\vF^H(t) =\mu_0\int_V \vJ(\vr,t)\times \vH(\vr,t) dV$.  However, this magnetic force is negligible compared to $\vF_{\pm Q}(t)$. To prove this, let the slowly decaying current have $\vJ_0(\vr)e^{-\alpha t}$ dependence for $\alpha\to 0$ so that $\vF_{\pm Q}(t)= O(\alpha)e^{-\alpha t}$.  Then the vector potential is
$$
\vA(\vr,t)\! = \!\frac{\mu_0}{4\pi}\!\int_V\!\! \frac{\vJ(\vr',t-R/c)}{R} dV'\! = \!e^{-\alpha t}\left[\frac{\mu_0}{4\pi}\!\int_V \!\frac{\vJ_0(\vr')}{R}e^{\alpha R/c} dV' \!=\! \frac{\mu_0}{4\pi}\!\int_V\! \frac{\vJ_0(\vr')}{R} dV' \!+\!\frac{\alpha\mu_0}{4\pi c}\!\int_V \!\vJ_0(\vr') dV' \!+\! O(\alpha^2)\! =\! \vA_0(\vr) \!+\! O(\alpha^2)\right]
$$
where $R=|\vr'-\vr|$ and we have made use of $\int_V\vJ_0(\vr)dV =0$ since $\nabla\cdot\vJ_0(\vr)=0$.  With $\vH(\vr,t)=\nabla\times\vA(\vr,t)/\mu_0$, we find $\vH(\vr,t)=[\vH_0(\vr) +O(\alpha^2)]e^{-\alpha t}$ and thus
$$
\vF^H(t) = e^{-\alpha t}\left[\mu_0\int_V\vJ_0(\vr)\times\vH_0(\vr)dV +O(\alpha^2)\right] = O(\alpha^2)e^{-\alpha t}
$$
because the magnetostatic self force $\mu_0\int_V\vJ_0(\vr)\times\vH_0(\vr)dV =0$ (which is easily proven by letting $V=V_\infty$, all space, and utilizing $\vJ_0(\vr)=\nabla\times\vH_0(\vr)$ along with $\nabla\cdot\vH_0(\vr)=0$ and standard vector identities).  Thus, $\vF^H(t)=O(\alpha^2)e^{-\alpha t}$ is negligible compared to $\vF_{\pm Q}(t)=O(\alpha)e^{-\alpha t}$ as $\alpha\to0$.}   Therefore, $\vG_Q$ in (\ref{S1}) is the total electromagnetic impulse imparted to the charge-current in the system %(disks and pillbox with radial arms and charges attached)
 as the angular velocity of each disk decays from its initial value to zero.
\par
The electromagnetic impulse in (\ref{S1}) is confirmed by evaluating the total Lorentz force expressed in terms of the Maxwell stress dyadic $\dT(\vr,t)$ [see (\ref{F76}) and (\ref{F90})] and the electromagnetic field momentum, $\vG_{el}=\int_V \vE\times\vH dV/c^2$ \cite[eq. (2.132)]{H&Y}
\be{S2}
\vF(t) = \int\limits_V[\rho(\vr,t)\vE(\vr,t)+\mu_0\vJ(\vr,t)\times\vH(\vr,t)]dV = \int\limits_S \vnh\cdot\dT(\vr,t)dS -\frac{1}{c^2}\frac{d}{dt}\int\limits_V \vE(\vr,t)\times\vH(\vr,t) dV
\ee
where the surface $S$ of the volume $V$ encloses all the charge and current.\footnote{The interpretation of $\vG_{el}=\int_V \vE\times\vH dV/c^2$ as electromagnetic field momentum follows from setting the total time rate of change of electromagnetic momentum, $\int_S \vnh\cdot\dT dS$ entering $S$ equal to the force $\vF$ on the charge-current within $S$ plus the time rate of change of the momentum of the electromagnetic fields within $S$.  If  $\int_S \vnh\cdot\dT dS=0$, then $\vF= -d\vG_{el}/dt$, that is, the force on the charge-current in $S$ equals the rate of decrease of the momentum of the fields in $S$.}  Letting $V$ be all space $V_\infty$, the surface integral of the stress dyadic on $S_\infty$ vanishes and because the radiated fields are negligible for our slowly varying system in Fig. \ref{Shockley}, (\ref{S2}) becomes, after integrating over the time interval $(0,\infty)$
\be{S3}
\int\limits_0^\infty\vF(t)dt = \frac{1}{c^2}\int\limits_{V_\infty} \vE(\vr,0)\times\vH(\vr,0) dV
\ee
where $\vE(\vr,0)=-\nabla\psi(\vr)$ is the electrostatic field of the two opposite point charges and $\vH(\vr,0)=\vH(\vr)$ is the initial magnetostatic field of the counter-rotating disks.  We can then write
\be{S4}
\int\limits_{V_\infty} \vE(\vr,0)\times\vH(\vr,0) dV = -\int\limits_{V_\infty}\nabla\psi(\vr)\times\vH(\vr)dV = 
\int\limits_{V_\infty}\psi(\vr)\nabla\times\vH(\vr)dV = \int\limits_V\psi(\vr)\vJ(\vr)dV
\ee
since $\int_{S_\infty}\vnh\times[\psi(\vr)\vH(\vr)]dS = 0$.  Because the current density $\vJ(\vr)$ is confined to the rotating disks that can be made arbitrarily small centered at the origin $\vr =0$, the scalar potential $\psi(\vr)$ of the two point charges can be approximated by the first term of its power series, that is
\be{S5}
\psi(\vr) = \psi(0)+\nabla\psi(0)\cdot\vr
\ee
and (\ref{S4}) becomes
\be{S6}
\int\limits_{V_\infty} \vE(\vr,0)\times\vH(\vr,0) dV =  \nabla\psi(0)\cdot\int\limits_V \vr\vJ(\vr)dV = -\vE_0\cdot\int\limits_V \vr\vJ(\vr)dV
\ee
where we have used $\int_V\vJ(\vr)=0$ to eliminate the first term from (\ref{S5}) in (\ref{S6}).
Rewriting $\vr\vJ = (\vr\vJ +\vJ\vr)/2 + (\vr\vJ -\vJ\vr)/2$, the first term $(\vr\vJ +\vJ\vr)/2$ integrated in (\ref{S6}) gives the electric quadrupole moment of the current distribution, which is zero, leaving
\be{S7}
\int\limits_{V_\infty} \vE(\vr,0)\times\vH(\vr,0) dV =  -\vE_0\cdot\frac{1}{2}\int\limits_V[\vr\vJ(\vr) -\vJ(\vr)\vr] dV = \vE_0\times\frac{1}{2}\int\limits_V[\vr\times\vJ(\vr)]dV = -\vm(0)\times\vE_0.
\ee
This result in (\ref{S7}), which was derived in a similar fashion by Calkin \cite{Calkin} and Vaidman \cite{Vaidman}, confirms that as the angular velocity of the disks decays to zero, the impulse in (\ref{S1}) indeed equals the initial electromagnetic field momentum, namely
\be{S8}
\vG_Q = \int\limits_0^\infty \vF_{\pm Q}(t)dt = \int\limits_0^\infty \vF(t)dt = \frac{1}{c^2}\int\limits_{V_\infty} \vE(\vr,0)\times\vH(\vr,0) dV = -\frac{1}{c^2}\vm(0)\times\vE_0.
\ee
\subsection{Replacing the Amperian magnetic dipole of the rotating-disks with a magnetic-charge magnetic dipole}\lbl{md}
If the counter-rotating disks creating an Amperian magnetic dipole moment are replaced by a magnetic-charge magnetic dipole formed by two equal and opposite magnetic charges to give a magnetic dipole moment $\vm$, the force from the electric-field of the time-varying magnetic dipole on the $\pm Q$ electric charges remains the same as the dipole moment slowly decays (for example, by letting the attracting opposite magnetic charges slowly slide together on a rod with friction) from a value of $\vm(0)$ at $t=0$ to a value of zero at $t=\infty$.  Now, however, the magnetic-charge magnetic dipole experiences a force as its magnetic charges slowly move in the electric fields of the $\pm Q$ charges.  This force is easily determined from the $-\eps_0 q_m\vv\times\vE_0$ forces on the moving magnetic charges to yield $-d\vm/dt\times \vE_0/c^2$.  Integrating this force over the time interval $(0,\infty)$, one obtains the electromagnetic impulse $\vG_m$ for the magnetic-charge dipole, namely
\be{S8'}
\vG_m = \int\limits_0^\infty \vF_m(t)dt =  \frac{1}{c^2}\vm(0)\times\vE_0.
\ee
{\color{black}Also, the equal and opposite forces between the two equal and opposite magnetic charges cancel.}
Therefore, the total force and momentum change ($\vG_{Qm}$) of the magnetic-charge dipole and electric charges is zero, that is
\be{S8''}
\vG_{Qm} = \vG_Q+\vG_m = 0.
\ee
\par
Again the total electromagnetic impulse in (\ref{S8''}) can be confirmed by evaluating the total Lorentz force expressed in terms of the Maxwell stress dyadic $\dT(\vr,t)$ and the electromagnetic field momentum, but now the Lorentz force and Maxwell's equations must include magnetic charge and current
\be{S2'}
\vF(t) = \int\limits_V[\rho\vE+\mu_0\vJ\times\vH + \rho_m\vH-\eps_0\vJ_m\times\vE]dV = \int\limits_S \vnh\cdot\dT dS -\frac{1}{c^2}\frac{d}{dt}\int\limits_V \vE\times\vH dV.
\ee
Then, we have as in (\ref{S3}),
\be{S3'}
\int\limits_0^\infty\vF(t)dt = \frac{1}{c^2}\int\limits_{V_\infty} \vE(\vr,0)\times\vH(\vr,0) dV.
\ee
Now, however $\nabla\times\vH(\vr)=0$ because the static magnetic field of the magnetic-charge magnetic dipole has no electric current and $\vH(\vr) = -\nabla\psi_m(\vr)$ to give
\be{S4'}
\int\limits_{V_\infty} \vE(\vr,0)\times\vH(\vr,0) dV = -\int\limits_{V_\infty}\vE(\vr)\times\nabla\psi_m(\vr)dV = 
-\int\limits_{V_\infty}\psi_m(\vr)\nabla\times\vE(\vr)dV = 0
\ee
since $\nabla\times\vE =0$ for the static electric field of the $\pm Q$ charges.
This result in (\ref{S4'}) shows that the zero total electromagnetic impulse in (\ref{S8''}) is indeed confirmed by a zero change in total electromagnetic momentum as the initial magnetic-charge magnetic dipole slowly decays to zero; in summary
\be{S8'''}
\vG_{Qm} = \vG_Q + \vG_m = \int\limits_0^\infty [\vF_{\pm Q}(t)+\vF_m(t)]dt = \int\limits_0^\infty \vF(t)dt  =\frac{1}{c^2}\int\limits_{V_\infty} \vE(\vr,0)\times\vH(\vr,0) dV = 0.
\ee
\par
Obviously, the total electromagnetic field momentum in (\ref{S8}) for the Amperian magnetic dipole and static electric charges is not equal to the total electromagnetic field momentum in (\ref{S8'''}) for the magnetic-charge magnetic dipole and static electric charges.  Moreover, the difference is not due to the frictional forces because in either model the frictional forces are equal and opposite so as not to contribute to the momentum.
\subsection{Additional forces on the charges of the counter-rotating disks}\lbl{af}
There are binding forces between the material of each disk and their charges that keep the charges of each disk uniformly spaced as the disks counter-rotate in the external electric field.  Shockley and James \cite{S&J} argue that these other ``mechanical stresses and motions"%\footnote{These ``mechanical stresses and motions" could be produced by the microscopic charges of the molecules comprising the material of the disks.}
 within the disks produce a net force between the charge carriers of the current and the material of the disks (the net force being transferred to the pillbox) that is equal and opposite to the force on $\pm Q$ and that creates an additional relativistic momentum in the disk material that cancels the electromagnetic field momentum $\vG_Q$.  % because in the absence of any external forces the momentum of a closed system where radiation is negligible must remain constant.
  In other words, these mechanical stresses and motions in the disks provide what Shockley and James call a ``hidden momentum" that bring the total force and linear momentum of the system with the Amperian magnetic dipole equal to the zero total force and linear momentum of the system with the magnetic-charge magnetic dipole.   %The charges on the disks exert opposite forces on the disk material and the $\pm Q$ charges on the radial arms exert opposite forces on the arms, giving a total change in momentum of the material equal to zero.  Thus, the total force and momentum-change of all the charges and material of the disks and pillbox with radial arms and charges attached is zero. 
\par
Coleman and Van Vleck \cite{C&V} substantiate the argument of Shockley and James \cite{S&J} by calling attention to the general theorem that any closed system with no external forces and no radiation that is described by a Lorentz-invariant Lagrangian has a stress-energy-momentum tensor that is symmetric and conserved.  Consequently, total momentum and energy of the system remain constant and the velocity of the center of energy is equal to the ratio of the total momentum to energy (times $c^2$).  Since before the disks begin to slow down, the center of energy does not change, it follows that the constant total momentum is zero for all time --- in particular, during the time that the angular velocities of the disks decrease.   %Notwithstanding the elegance of this theorem, one could question whether it is any more fundamental or appealing to assume a Lorentz-invariant Lagrangian than Newton's third law of motion for a physical system in which time delays and radiation are negligible.
\par
There is little doubt that Shockley and James as well as Coleman and Van Vleck are correct in concluding that a quasistatic Amperian magnetic dipole in a static external field must contain a so-called ``hidden momentum" if radiation is negligible such that the quasistatic Amperian magnetic dipole experiences the same force as an equivalent quasistatic magnetic-charge magnetic dipole. %(see Appendix \ref{AppendixA})
 Still one can ask if the exact same value of ``hidden momentum" results for arbitrarily time-varying Amperian magnetic dipoles that radiate in arbitrarily time-varying external fields;  moreover, exactly how does this hidden momentum manifest itself. %and is it truly ``mechanical" in nature or is it entirely or partly electromagnetic in nature. For example, assuming the Amperian magnetic dipole can be modeled by nonradiating electrons orbiting an oppositely charged nucleus with the nucleus attached to a rigid lattice, how much of the hidden-momentum force changes the kinetic momentum of the orbiting electrons and how much of it is transferred to the rigid lattice as a ``structural force", that is, the force exerted on the rigid structure which holds the nucleus of the Amperian dipole in place.\footnote{The rigid lattice can translate and rotate (like the disks) without changing the structural force as long as the acceleration is such that the net force between the nuclei and the lattice is changed negligibly.}
  Since the appearance of the original two papers, \cite{S&J} and \cite{C&V},  these questions (and others like them) have been the subject of many publications proposing several different models for the Amperian magnetic dipole and leading to different conclusions that are sometimes conflicting.  We will briefly review a few of the most important classical models of Amperian magnetic dipoles in order to explain the main motivation for the present paper.
{\color{black}
\subsection{Classical models of Amperian magnetic dipoles}\lbl{Cm}}
At about the same time that Shockley and James published their paper, Penfield and Haus \cite[sec. 7.4]{P&H} determined that charged particles flowing around a loop in a frictionless tube (to create {\color{black}a uniform-current} Amperian magnetic dipole moment $\vm$) and subject to a uniform constant external electric field $\vE_0$ would possess a net kinetic momentum equal to $\vm\times\vE_0/c^2$ produced by the variation in each of the charged particle's relativistic kinetic energy induced by the external field $\vE_0$ as the charged particles travel around the loop; see also Griffiths \cite[pp. 547--549]{Griffiths}.
\par
If this tube model of the Penfield-Haus Amperian magnetic dipole replaces the current of the disks in the Shockley-James system, then as the circulating charges in the tube slow down because of a slight amount of friction in the tube, the electric field of the changing magnetic dipole moment will impart the same impulse $\vG_Q$ to the $\pm Q $ charges.  During this slowdown, the force on the $\pm Q$ charges will be $d\vm/dt\times\vE_0/c^2$.  The forces between the circulating charges in the tube and the $\pm Q$ will simply be the equal and opposite Coulomb forces because the Penfield-Haus model assumes uniform solenoidal current ($\nabla\cdot\vJ = -\partial \rho/\partial t =0$).  Nonetheless, there has to be an additional force on the circulating charges because their kinetic momentum changes as  $d\vm/dt\times\vE_0/c^2$.  This ``hidden-momentum" force on the charges circulating inside the tube can be supplied by the radial and frictional tangential tube forces or by the mutual forces between the circulating charges with changing velocities (changing internal electromagnetic momentum).  The Penfield-Haus model does not answer the question of how the required hidden-momentum force divides between the tube forces and internal electromagnetic forces because it ignores the forces exerted on the charges by their mutual fields as well as the radiation reaction force on each charge.  Also, the Penfield-Haus model is limited to a static external electric field and a static or, at best, a quasistatic magnetic dipole.  Overall, it is a clever but inadequate classical model for microscopic molecular time-varying Amperian magnetic dipoles in time varying external fields.  
\par
Boyer \cite{Boyer}, \cite{BoyerAJP} improves upon the model of Penfield and Haus by considering a magnet composed of any number of interacting charges circulating in a circular ``ring" path (corresponding to the Penfield-Haus tube) with a compensating opposite charge on a particle fixed at the center of the circular ring, all subject to the static external electric field of a distant point charge.  External forces tangential to the circular ring are then applied to the circulating charges that quasistatically change the magnetic dipole moment linearly with time.  These tangential forces correspond to the frictional forces in the Shockley-James counter-rotating disks and to the frictional forces in the Penfield-Haus tube.  Boyer also considers the radial external forces needed to constrain the circulating charges to the circular ring path.  Thus, Boyer's total external force on the circulating charges corresponds to the total force exerted by the tube on the circulating charges in the Penfield-Haus model.  Denoting this total force by $\vF_{\rm tube}(t)$, Newton's second law of motion applied to the circulating charges demands that  
\be{Boyer1}
\vF_{\rm tube}(t) +\vF^c_{el}(t) = \frac{d\vG^c_K(t)}{dt}
\ee
where $\vF^c_{el}(t)$ is the net force between the circulating charges, which can be written in terms of the internal electromagnetic momentum of the fields of the circulating charges as $\vF^c_{el}(t)=-d\vG^c_{el}(t)/dt$, and $\vG^c_K(t)$ is the kinetic momentum of the circulating charges (referred to as ``mechanical momentum" by Boyer).  The force equation in (\ref{Boyer1}) can be rewritten as
\be{Boyer2}
\vF_{\rm tube}(t)  = \frac{d\vG^c_{el}(t)}{dt} + \frac{d\vG^c_K(t)}{dt}.
\ee
For a single circulating charged particle, $d\vG^c_{el}/dt =0$ and $\vF_{\rm tube} = d\vG^c_K/dt = d\vm/dt\times \vE_0/c^2$ when averaged over successive cycles.  For two charged particles interacting at low velocities, Boyer finds that the kinetic momentum is negligible ($d\vG^c_K/dt \approx 0$) and that $\vF_{\rm tube} \approx d\vG^c_{el}/dt \approx d\vm/dt\times \vE_0/c^2$.  In either case, the total time rate of change of momentum of the circulating charges is equal to the ``hidden-momentum" force, $d\vm/dt\times \vE_0/c^2$, which equals $d(\vm\times\vE_0)/dt/c^2$ because Boyer's $\vE_0$ is static (although Boyer discourages the use of the term ``hidden momentum").  Boyer assumes (without proof) that for a finite number of charged particles interacting with unrestricted velocities, the sum of the time rate of change of the kinetic and internal electromagnetic momentum (equal to $\vF_{\rm tube}(t)$) will always equal $d(\vm\times\vE_0)/dt/c^2$ and, as the number of charged particles approaches infinity, the time rate of change of the kinetic momentum will approach zero.  (It should be noted that the sum of the electrostatic Coulomb forces on Boyer's neutral magnetic dipole model is zero except for the electric  dipole force $\vp_0\cdot\nabla\vE_0$ on the electric dipole moment $\vp_0$ that is induced in the circulating charges by the external field $\vE_0$.  However, this well-known electric dipolar force does not change the hidden momentum in the Amperian magnetic dipole and Boyer assumes $\nabla\vE_0\approx0$ so that in his case the electric dipolar force is negligible.)
\par
The shortcomings of Boyer's illuminating analysis are that it neglects terms in the particle velocities higher than second order, that it ignores radiation reaction forces on each of the charged particles, and that it is limited to static external electric fields and quasistatic Amperian magnetic dipoles changing linearly with time.
\par
In an often overlooked, though impressive derivation, de Groot and Suttorp, in their book on the ``Foundations of Electrodynamics" \cite[pp. 195--196]{D&S}, consider classical Amperian models of ``stable atoms" consisting of a number of interacting charged particles in an external electromagnetic field.  They find that there is indeed a term equal to the hidden momentum and that the time rate of change of this term is manifested as a force accelerating the center of energy of the composite stable atom.\footnote{De Groot and Suttorp find an angular momentum term added to the $\vm$ in the hidden momentum.  However, this term is negligible for electrons whose mass is small compared to the mass of the nucleus of the atom.}  However, for ``stable atoms" consisting of electrons circulating a nucleus fixed to a rigid lattice, the analyses in \cite{D&S} do not reveal how much of the hidden-momentum of the ``stable atom" is kinetic and how much is internal electromagnetic.  Also, since the de Groot-Suttorp model for classical ``stable atoms" ignores the radiation reaction forces on the charged particles (effectively, the charged particles are not allowed to radiate),  this model of ``stable atoms" in an external electromagnetic field does not provide an exact classical solution for the internal momentum and force.   Indeed, all exact classical solutions to charged particles circulating an oppositely charged nucleus in free space are unstable.
\par
Probably the simplest, most appealing, rigorously tractable stable classical model for the electric and magnetic dipole moments of molecules are electrically small perfect electric conductors (PEC's), where the term PEC is used here in the sense of a superconductor whose internal electric and magnetic fields are zero even in the case of static fields. Both electric and magnetic dipole moments can be induced by externally applied fields on a singly connected PEC and a stable static magnetic dipole moment can exist on a doubly connected PEC without an externally applied field, for example, on a wire loop.  (Notably, Weber and Maxwell \cite[arts. 836--845]{Maxwell} explained both diamagnetism and ordinary magnetism (paramagnetism or ferro(i)magnetism) by means of PEC wire loops with no initial static current in the case of diamagnetism, and predominantly initial static current in the case of  ordinary magnetism \cite{Yaghjian-Reflection}, \cite{Yaghjian-Classical}.)  A stable static electric dipole moment can be modeled by equal and opposite electric charges on two PEC's separated by an insulating rod.
\par
As far as I am aware, no general proof exists for the forces on electrically small PEC's in a time-varying external field and, in particular, for deciding unequivocally whether such Amperian magnetic dipoles give rise to a hidden-momentum force that makes the force they experience in an external field identical to the force that a magnetic-charge magnetic dipole with the same dipole moment $\vm(t)$ would experience in the same external field.  Indeed, Vaidman \cite{Vaidman} challenged his readers ``to provide a proof [of the hidden momentum] for an arbitrary shape of a conductor [PEC]."  Hnizdo \cite{Hnizdo}, using ideas from Furry \cite{Furry}, responded to Vaidman's challenge by dividing the internal force induced by the external field on the conducting loop into two contributions, one from a static magnetic dipole moment in a time-varying external electric field and another from a time-varying magnetic dipole moment in a static external electric field.  The first contribution further divides into two parts:  the force exerted by the magnetic field of the moving induced charges $q_i$, each assigned a velocity $\vv_i$, on a solenoidal current with constant $\vm$ plus the force exerted by the magnetic field of the constant solenoidal current on the moving induced charges --- the sum yielding a force $- \vm\times d\vE/dt/c^2$, where $\vE$ is the time-varying externally applied electric field.  The second contribution is equal to the force exerted on the static charge (induced by a static external electric field) by the time-varying electric field from the vector potential of the time-varying magnetic dipole moment  --- yielding a force $- d\vm/dt\times \vE/c^2$.  The sum of the two contributions gives the hidden-momentum force $-d(\vm\times\vE)/dt/c^2$.
\par
Notwithstanding the appeal of Hnizdo's analysis, it nonetheless involves several approximations (in addition to the initial assumption that the force can be separated into two uncoupled contributions).  The velocity $\vv_i$ is not an actual physical velocity of the induced charge (which can be positive or negative) because only the negative electrons move in the conductor.  Even if we grant $\vv_i$ as an effective electron velocity, Hnizdo's derivation approximates the electric field of a moving charge in its instantaneous rest frame by the Coulomb field while ignoring the rest-frame magnetic field (caused by $\dot{\vv}_i$).  His derivation neglects both the force exerted by the magnetic field of each moving induced charge on the other moving induced charges and the force exerted by the time-varying magnetic field of the solenoidal part of the current on the solenoidal current itself, as well as the force exerted on the electric charges by the electric field of the time-varying induced scalar potential.  Proof is needed that these approximations are valid and that the neglected contributions are indeed negligible.\footnote{If a circular PEC wire loop replaces the counter-rotating disks in the Shockley-James system, the static electric field $\vE(\vr,0)$ in (\ref{S4}) becomes the sum of the electric field of the external static charges ($\pm Q$) and the electric field induced on the PEC wire loop.  This sum total electric field has to have a zero tangential component across the surface of the PEC wire loop so that $\psi(\vr)$ is a constant $\psi_0$ in the last integral of (\ref{S4}).  Thus, this integral is zero ($\psi_0\int_V \vJ(\vr)dV = 0$) and, like the total momentum for the static magnetic-charge magnetic dipole in an external static electric field (see (\ref{S4'})),  the total momentum for the static PEC Amperian magnetic dipole in an external static electric field is zero, as Calkin \cite{Calkin} and Vaidman \cite{Vaidman} have noted previously.  Nonetheless, this simple proof relies upon the sources and fields being static and does not apply to general time-varying PEC Amperian magnetic dipoles in time-varying external fields.}
\par
Essentially, all these previous derivations of hidden momentum are at best implicitly or explicitly restricted to slowly varying radiationless dipoles subject to static or quasistatic external fields and, thus, they merely verify the quasistatic radiationless conservation theorems originally used by Coleman and Van Vleck \cite{C&V} to prove the existence of hidden momentum under these restrictions.   Consequently, the main purpose of the present paper is to rigorously derive from Maxwell's equations the dipolar forces on an arbitrarily shaped electrically small %stationary (fixed in the laboratory reference frame)
 radiating PEC in a {\color{black}bandlimited but otherwise} general time-varying external electromagnetic field.  In particular, the existence of electromagnetic hidden momentum is verified that ensures that the forces on time-varying PEC-Amperian and magnetic-charge magnetic dipoles in time-varying external fields are equal.  Moreover, because the tangential electric field across the surface of a %stationary 
PEC is zero, there is no variation in the kinetic energy of the charges around the solenoidal current paths and, therefore, no net kinetic momentum is induced in the PEC-Amperian magnetic dipoles.  Thus, the hidden-momentum force is an internal electromagnetic force on the charge-current of the PEC that is transferred as a structural force exerted on whatever frame holds the PEC fixed.
\par
The exact Mie solution for scattering of a plane wave by an electrically small PEC rigid sphere is used to confirm the general expressions for the forces obtained for the time varying electric and magnetic dipoles of an arbitrarily shaped electrically small PEC in time varying external fields.  
%\par
%Lastly, a general argument that requires relatively little mathematics is given, based on Newton's third law of equal and opposite forces applied to charge carriers interacting in each others quasistatic fields, to prove that hidden momentum is always contained in electrically small Amperian magnetic dipoles to make equal the total force on passive Amperian and magnetic-charge magnetic dipoles in time-varying external fields, whether or not the magnetic dipoles are lossy or lossless.  The Mie solution for an electrically small sphere with finite conductivity is used to further demonstrate that the same expression for hidden momentum holds for lossy and lossless Amperian magnetic dipoles.
\par
We use the adjective ``microscopic" in the title of the paper and throughout to indicate that we are not treating models of dipoles that involve materials characterized by bulk parameters because such materials, for example, dielectric or magnetic materials characterized by an $\eps\neq\eps_0$ or $\mu\neq\mu_0$, require macroscopic averaging of microscopic dipolar fields and forces that could result in expressions for the macroscopic forces in terms of macroscopic dipole moments and fields that differ from the microscopic expressions.
\section{FORCE ON ELECTRIC AND MAGNETIC DIPOLES}\label{FEMD}
We want to determine from first principles the force experienced by the total charge and current densities $[\rho(\vr,t),\vJ(\vr,t)]$ on an electrically small rigid %stationary 
PEC situated in free space within a circumscribing sphere of radius $a$ and illuminated by externally applied electromagnetic fields $[\vE_e(\vr,t),\vH_e(\vr,t)]$.   By ``electrically small", we mean that the three-dimensional spatial Fourier transform of the external fields is bandlimited to a value $k_{\rm max}=|\vk|_{\rm max}$ such that $k_{\rm max}a\ll 1$, where $\vk$ is the three-dimensional vector Fourier spatial frequency.  These spatially bandlimited external fields induce charge and current densities $[\rho_s(\vr,t),\vJ_s(\vr,t)]$ which, in turn, produce electric and magnetic dipole moments $[\vp_s(t),\vm_s(t)$] but negligible higher-order spherical multipole moments (but not necessarily negligible higher-order multipolar surface fields) for $k_{\rm max}a\ll1$.  The PEC is allowed to carry static charge and current densities $[\rho_0(\vr),\vJ_0(\vr)]$ producing the static electric and magnetic dipole moments $[\vp_0,\vm_0$].  The induced charge and current produce the scattered fields $[\vE_s(\vr,t),\vH_s(\vr,t)]$ everywhere and the static charge and current produce the static fields $[\vE_0(\vr),\vH_0(\vr)]$ everywhere, so that
\begin{subequations}
\lbl{F1}
\be{F1a}
\rho(\vr,t) = \rho_s(\vr,t) + \rho_0(\vr)
\ee
\be{F1b}
\vJ(\vr,t) = \vJ_s(\vr,t) + \vJ_0(\vr)
\ee
\be{F1c}
\vp(t) = \vp_s(t) + \vp_0
\ee
\be{F1d}
\vm(t) = \vm_s(t) + \vm_0
\ee
\be{F1e}
\vE(\vr,t) = \vE_e(\vr,t) + \vE_s(\vr,t) + \vE_0(\vr)
\ee
\be{F1f}
\vH(\vr,t) = \vH_e(\vr,t) + \vH_s(\vr,t) + \vH_0(\vr).
\ee
\end{subequations}
%
%The sources of the externally applied fields lie outside the radius $a$ of the PEC circumscribing sphere.
\par
The Lorentz force exerted by the fields on the charge and current is given by\footnote{The Lorentz force on $\rho dV$ and $\vJ dV$ is exerted by the fields of all sources except the fields of $\rho dV$ and $\vJ dV$ themselves. However, the self-fields of an infinitesimal volume element of continuous volume density of charge and current approach zero as $dV \to 0$ and thus $\vE$ and $\vH$ in the volume-element Lorentz force are the total electric and magnetic fields.  Also, since the charge carriers of each differential element of continuous charge and current, $\rho dV$ and $\vJ dV$, can be moving, they can experience a self force \cite{YaghjianBook}.  However, since the self force is proportional to $(\rho dV)^2$ and $|\vJ dV|^2$, it is a higher-order differential than $dV$ and thus the self force can be ignored in the volume integrals of (\ref{F7}). (This does not imply that the integrated self-force for a fixed amount of charge-current moving as a relativistically rigid charged particle is negligible \cite{YaghjianBook}.)}
\bea{F7} 
\vF(t) = \int\limits_V [\rho(\vr,t)\vE(\vr,t) + \mu_0\vJ(\vr,t)\times\vH(\vr,t)]dV \hspace{40mm}\\ = \int\limits_V [\rho\vE_e + \mu_0\vJ\times\vH_e]dV + \int\limits_V [\rho(\vE_s+\vE_0) + \mu_0\vJ\times(\vH_s+\vH_0)]dV\nonumber
\eea
where $V$ is any volume enclosing the charge and current [$\rho(\vr,t),\vJ(\vr,t)$].
\subsection{Force exerted directly by the external fields}\label{FEEF}
The force exerted directly by the external fields is given from (\ref{F7}) as
\be{F8}
\vF_e(t) = \int\limits_V [\rho(\vr,t)\vE_e(\vr,t) + \mu_0\vJ(\vr,t)\times\vH_e(\vr,t)]dV.
\ee
Since the sources of the external fields lie outside the PEC, the external fields can be expanded in a power series  about the center ($\vr=0$) of the sphere that circumscribes the electrically small PEC, namely
\begin{subequations}
\lbl{F9}
\be{F9a}
\vE_e(\vr,t) = \vE_e(0,t) + \vr\cdot\nabla\vE_e(0,t) + \ldots
\ee
\be{F9b}
\vH_e(\vr,t) = \vH_e(0,t) + \vr\cdot\nabla\vH_e(0,t) + \ldots
\ee
\end{subequations}
where only the first order terms in $\vr$ need be kept because the higher order $\vr$ terms lead to forces on spherical multipole moments of higher order than dipoles and we are assuming electrically small enough scatterers that these higher order multipole-moment forces are negligible in bandlimited external fields.
\par
Under the assumption that the total charge on the PEC is zero, that is
\be{F11}
\int\limits_V \rho(\vr,t)dV =0
\ee
the integral of $\vE_e(0,t)$ from (\ref{F9a}) inserted into (\ref{F8}) is zero.  This leaves the integral
\be{F12}
\vF^E_e(t) = \int\limits_V \rho(\vr,t)\vE_e(\vr,t)dV = \left[\int\limits_V \rho(\vr,t)\vr dV\right]\cdot\nabla\vE_e(0,t) = \vp(t)\cdot\nabla\vE_e(0,t)
\ee
since
\be{F12'} 
\vp(t) =\int\limits_V \vr\rho(\vr,t) dV.
\ee
With the help of the identity $\nabla\cdot(\vJ\vr) = (\nabla\cdot\vJ)\vr +\vJ$ and the definition of the electric dipole moment in (\ref{F12'}), we find
\be{F13}
\int\limits_V \vJ(\vr,t)dV = -\int\limits_V \nabla\cdot\vJ(\vr,t)\vr dV = \int\limits_V \frac{\partial\vrho(\vr,t)}{\partial t}\vr dV = \frac{d\vp(t)}{dt}
\ee
where we have used $\int_V \nabla\cdot (\vJ\vr)dV = \int_S \vnh\cdot(\vJ\vr)dS =0$ with $S$ being the surface of the enclosing volume $V$.
The result in (\ref{F13}) yields for the $\vH_e(0,t)$ term from (\ref{F9b}) substituted into the integral in (\ref{F8})
\be{F14}
\mu_0 \frac{d\vp(t)}{dt}\times\vH_e(0,t).
\ee
The remaining integral obtained from (\ref{F9b}) inserted into (\ref{F8}) is
\be{F15}
\int\limits_V  \vJ(\vr,t)\times[\vr\cdot\nabla\vH_e(0,t)]dV.
\ee
To evaluate this integral, use the identities
\begin{subequations}
\lbl{F16}
\be{F16a}
\vJ\times(\vr\cdot\nabla\vH_e) = \nabla H_{ex} \cdot(\vr\vJ)\times\vxh + \nabla H_{ey} \cdot(\vr\vJ)\times\vyh + \nabla H_{ez} \cdot(\vr\vJ)\times\vzh
\ee
\be{F16b}
\vr\vJ = \frac{1}{2}(\vr\vJ + \vJ\vr) + \frac{1}{2}(\vr\vJ - \vJ\vr) 
\ee
\be{F16c}
\va\cdot(\vr\vJ - \vJ\vr) =  (\vr\times\vJ)\times\va
\ee
\end{subequations}
and the definition of the magnetic dipole moment
\be{F18'}
\vm(t) = \frac{1}{2}\int\limits_V \vr\times\vJ(\vr,t) dV
\ee
to get
\be{F19}
\int\limits_V  \vJ(\vr,t)\times[\vr\cdot\nabla\vH_e(0,t)]dV = (\vm\times\nabla H_{ex})\times \vxh + (\vm\times\nabla H_{ey})\times \vyh + (\vm\times\nabla H_{ez})\times \vzh
\ee
where we have dropped the $\int_V (\vr\vJ + \vJ\vr)dV$ term, which can be shown to equal $d/dt\int_V \rho \vr\vr dV = d\dQ/dt$, the time derivative of the electric quadrupolar moment, which is negligible compared with the electric and magnetic dipole moments of electrically small enough PEC's.  The expression in (\ref{F19}) can be rewritten as
\be{F20}
\int\limits_V  \vJ(\vr,t)\times[\vr\cdot\nabla\vH_e(0,t)]dV = \nabla\vH_e\cdot\vm = \vm \cdot \nabla\vH_e + \vm\times\nabla\times\vH_e = \vm \cdot \nabla\vH_e + \eps_0\vm\times\frac{\partial\vE_e}{\partial t}
\ee
because $\nabla\cdot\vH_e = 0$ and from Maxwell's first equation for the externally applied fields, $\nabla\times\vH_e = \eps_0\partial \vE_e/\partial t$.  Adding (\ref{F20}) to (\ref{F14}) gives
\be{F20'}
\vF_e^H(t) = \int\limits_V \mu_0\vJ(\vr,t)\times\vH(\vr,t)dV = \mu_0 \frac{d\vp(t)}{dt}\times\vH_e(0,t) + \mu_0\vm(t)\cdot\nabla\vH_e(0,t) + \mu_0\eps_0\vm(t)\times\frac{\partial\vE_e(0,t)}{\partial t}.
\ee
In all then, we have evaluated $\vF_e(t)$ in (\ref{F8}) for electric and magnetic dipole moments as
\be{F21}
\vF_e(t) = \vF^E_e(t)+\vF^H_e(t) = \vp(t)\cdot\nabla\vE_e(0,t) + \mu_0 \frac{d\vp(t)}{dt}\times\vH_e(0,t) + \mu_0\vm(t)\cdot\nabla\vH_e(0,t) + \mu_0\eps_0\vm(t)\times\frac{\partial\vE_e(0,t)}{\partial t}.
\ee
The first and third terms on the right-hand side of (\ref{F21}) are the well-known quasistatic forces exerted by the external electric and magnetic fields on the electric and magnetic dipoles, respectively.  The second term is an understandable magnetic force exerted by the external magnetic field on the equivalent electric current of the time-varying electric dipole moment $d\vp/dt$.  The last term on the right-hand side of (\ref{F21}), $\mu_0\eps_0\vm\times \partial\vE_e/\partial t$, is not so easy to explain.  And, in fact, we shall show that an evaluation of the internal forces on the Amperian magnetic dipole produces a ``hidden-momentum" force which adds to this term to yield a total force equal to the force exerted on an equivalent magnetic-charge magnetic dipole by the external electric field.
\subsection{Force exerted by the internal fields}\label{FEIF}
The force exerted by the internal fields on the charge and current densities of the PEC is given by the last volume integral in (\ref{F7}), namely
\bea{F22}
\vF_{i}(t) =  \int\limits_V \{\rho(\vr,t)[\vE_s(\vr,t)+\vE_0(\vr)] + \mu_0\vJ(\vr,t)\times[\vH_s(\vr,t)+\vH_0(\vr)]\}dV\nonumber\\ = \int\limits_V [\rho(\vr,t)\vE_i(\vr,t) + \mu_0\vJ(\vr,t)\times\vH_i(\vr,t)]dV \hspace{30mm} 
\eea
where
\begin{subequations}
\lbl{F23}
\be{F23a}
\vE_i(\vr,t) = \vE_s(\vr,t) + \vE_0(\vr)
\ee
\be{F23b}
\vH_i(\vr,t) =  \vH_s(\vr,t) + \vH_0(\vr).
\ee
\end{subequations}
The internal fields satisfy the Maxwell equations
\begin{subequations}
\lbl{F25}
\be{F25a}
\nabla\times\vE_i(\vr,t) +\mu_0\frac{\partial\vH_i(\vr,t)}{\partial t} = 0
\ee
\be{F25b}
\nabla\times\vH_i(\vr,t) -\eps_0\frac{\partial\vE_i(\vr,t)}{\partial t} = \vJ(\vr,t)
\ee
\be{F25c}
\nabla\cdot\vE_i(\vr,t) = \frac{\rho(\vr,t)}{\eps_0}
\ee
\be{F25d}
\nabla\cdot\vH_i(\vr,t) = 0
\ee
with the continuity equation following from (\ref{F25b}) and (\ref{F25c})
\be{F25e}
\nabla\cdot\vJ(\vr,t) = -\frac{\partial \rho(\vr,t)}{\partial t}.
\ee
\end{subequations}
\par
We want to use these Maxwell equations to evaluate the force $\vF_i(t)$ in (\ref{F22}).  Toward this end, we divide the current and charge densities into\footnote{This decomposition of $\vJ$ into $\vJ_1+\vJ_2$ is not the familiar Helmholtz decomposition into solenoidal and irrotational vectors \cite[secs. 83--89]{Phillips} because even though $\vJ_2$ is solenoidal, $\vJ_1$ is not irrotational and thus not equal to the gradient of a scalar function.  Also, unlike the Helmholtz decomposition, our $\vJ_1$ and $\vJ_2$ have compact support (are zero outside the volume of the PEC).}
\begin{subequations}
\lbl{J12}
\be{J12a}
\vJ(\vr,t)= \vJ_1(\vr,t) + \vJ_2(\vr,t),\;\;\;[\vJ_1(\vr,t)=0, \vJ_2(\vr,t)=0]\mbox{ for } \vr \notin \mbox{PEC}
\ee
\be{J12b}
\rho(\vr,t)= \rho_1(\vr,t) + \rho_2(\vr,t),\;\;\;[\rho_1(\vr,t)=0, \rho_2(\vr,t)=0]\mbox{ for } \vr \notin \mbox{PEC}
\ee
\end{subequations}
with
\begin{subequations}
\lbl{F26}
\be{F26a}
\nabla\cdot\vJ_2(\vr,t) = -\frac{\partial\rho_2(\vr,t)}{\partial t} = 0
\ee
\be{F26b}
\nabla\cdot\vJ_1(\vr,t) = \nabla\cdot\vJ(\vr,t) - \nabla\cdot\vJ_2(\vr,t) = -\frac{\partial\rho_1(\vr,t)}{\partial t} = -\frac{\partial\rho(\vr,t)}{\partial t}
\ee
\end{subequations}
which imply that
\begin{subequations}
\lbl{F27}
\be{F27a}
\rho_2(\vr,t) = -\rho_0(\vr),\;\;\; \rho_0(\vr)=0\mbox{ for } \vr \notin \mbox{PEC}
\ee
is a time independent charge density and
\be{F27b}
\rho_1(\vr,t) = \rho(\vr,t) + \rho_0(\vr).
\ee
\end{subequations}
It can be proven \cite[sec. 1.10 and app. A]{R&D} that the solenoidal current density $\vJ_2(\vr,t)$ in the PEC produces only magnetic multipole moments.  Since the solenoidal current $\vJ_2$ in (\ref{J12a}) is not uniquely determined by the condition that $\nabla\cdot\vJ_2=0$, we can further specify that {\color{black} the magnetic multipole moments are produced by $\vJ_2$ and, in particular,} the magnetic dipole moment of $\vJ_2$ equals the dipole moment $\vm$ of the PEC, that is
\be{F26-m}
\vm(t) = \frac{1}{2}\int\limits_V \vr\times\vJ_2(\vr,t) dV.
\ee
The remaining charge-current density $[\rho(\vr,t),\vJ_1(\vr,t)]$ in the PEC produces the electric multipole moments and, in particular, the electric dipole moment $\vp$ of the PEC.  The charge density $-\rho_0(\vr)$ in $\vJ_2$ and the opposite charge density $\rho_0(\vr)$ in $\vJ_1$ cancel to contribute zero net fields.  In a homogeneous conductor, $-\rho_0$ and $\rho_0$ are the uniform (independent of $\vr$) cancelling time-independent electron and positive-charge densities, respectively, throughout the conductor.  
\par
The solenoidal current $\vJ_2(\vr,t)$ can be expressed in terms of its associated charge density $\rho_2=-\rho_0(\vr)$ times the velocity $\vv_2(\vr,t)$ of $\rho_2$, that is
\begin{subequations}
\lbl{F26'}
\be{F26'a}
\vJ_2(\vr,t) = -\rho_0(\vr)\vv_2(\vr,t)
\ee
so that
\be{F26'b}
\vJ_1(\vr,t) = \vJ(\vr,t) + \rho_0(\vr)\vv_2(\vr,t)
\ee
\end{subequations}
although this representation of $\vJ_2$ is not required for the subsequent derivations.  As the conductivity approaches an infinite value, the conductor becomes a perfect electric conductor with $\rho_0$ approaching infinity and $\rho_0\vv_2(\vr,t)$ approaching zero for all $\vr$ except near the surface of the conductor such that $-\rho_0\vv_2(\vr,t)=\vJ_2(\vr,t)$ approaches a surface delta function.  A type-1 superconductor, which has a finite electron charge density $-\rho_0$ that moves frictionlessly through the stationary positive charge density $\rho_0$, approximates this behavior with $\vJ_2=-\rho_0\vv_2(\vr,t)$ (and the accompanying interior magnetic field) confined to a thin surface layer that, nonetheless, covers many interatomic distances \cite[ch. 6]{LLP}. 
\par
The two charge-current densities produce their own internal electromagnetic fields obeying the Maxwell equations
\begin{subequations}
\lbl{F28}
\be{F28a}
\nabla\times\vE_{i1}(\vr,t) +\mu_0\frac{\partial\vH_{i1}(\vr,t)}{\partial t} = 0
\ee
\be{F28b}
\nabla\times\vH_{i1}(\vr,t) -\eps_0\frac{\partial\vE_{i1}(\vr,t)}{\partial t} = \vJ_1(\vr,t)
\ee
\be{F28c}
\nabla\cdot\vE_{i1}(\vr,t) = \frac{\rho(\vr,t)}{\eps_0}
\ee
\be{F28d}
\nabla\cdot\vH_{i1}(\vr,t) = 0
\ee
\end{subequations}
and
\begin{subequations}
\lbl{F29}
\be{F29a}
\nabla\times\vE_{i2}(\vr,t) +\mu_0\frac{\partial\vH_{i2}(\vr,t)}{\partial t} = 0
\ee
\be{F29b}
\nabla\times\vH_{i2}(\vr,t) -\eps_0\frac{\partial\vE_{i2}(\vr,t)}{\partial t} = \vJ_2(\vr,t)
\ee
\be{F29c}
\nabla\cdot\vE_{i2}(\vr,t) = 0
\ee
\be{F29d}
\nabla\cdot\vH_{i2}(\vr,t) = 0
\ee
\end{subequations}
where 
\begin{subequations}
\be{F30a}
\vE_i(\vr,t) = \vE_{i1}(\vr,t) + \vE_{i2}(\vr,t)
\ee
\be{F30b}
\vH_i(\vr,t) = \vH_{i1}(\vr,t) + \vH_{i2}(\vr,t).
\ee
\end{subequations}
The $\rho_0(\vr)$ on the right-hand side of (\ref{F28c}) and the $-\rho_0(\vr)$ on the right-hand side of (\ref{F29c}) have been omitted because they create equal and opposite electrostatic fields that cancel when $\vE_{i1}$ and $\vE_{i2}$ are added to give the total electric field $\vE_i$. In fact, both $\rho_0(\vr)$ and $-\rho_0(\vr)$ can be associated with $\vJ_2(\vr)$ such that $\rho_2(\vr)=0$; then $\rho_1(\vr)=\rho(\vr)$.
\subsubsection{Near fields of electrically small PEC scatterers}\label{Nearfields}
So far we have not made use of the electrically small size of the PEC in which the currents $\vJ_1(\vr,t)$ and  $\vJ_2(\vr,t)$ of the electric and magnetic dipole moments are induced by the spatially bandlimited externally applied fields.  As explained above, what is meant by ``electrically small" is that the PEC scatterer is illuminated by time-domain electromagnetic fields with $k_{\rm max}a \ll 1$ where $k_{\rm max}$ is the maximum significant magnitude of the vector Fourier spatial frequency in the operational bandwidth of the time-domain fields and $a$ is the radius of the sphere that circumscribes the PEC scatterer.  For external fields composed of a spectrum of propagating plane waves, $k_{\rm max} = \w_{\rm max}/c = 2\pi/\lambda_{\rm min}$, where $\w_{\rm max}$ is the maximum significant temporal angular frequency ($\lambda_{\rm min}$ the minimum wavelength) in the spectrum of the plane waves.
\par
The solution to Maxwell's equations in (\ref{F28}) and (\ref{F29}), Fourier transformed to the frequency domain ($e^{-i\w t}$ dependence), can be formally expressed in terms of the frequency-domain vector and scalar potentials \cite[sec. 1.9]{Stratton}, \cite[sec. 2.3.6]{H&Y}
\begin{subequations}
\lbl{F31}
\be{F31a}
\vE^\w_{i1}(\vr) = -\nabla\psi_{i1}^\w(\vr) +i\w \vA_{i1}^\w(\vr)
\ee
\be{F31b}
\vH^\w_{i1}(\vr) = \frac{1}{\mu_o}\nabla\times \vA_{i1}^\w(\vr)
\ee
\end{subequations}
with
\begin{subequations}
\lbl{F32}
\be{F32a}
\psi^\w_{i1}(\vr) = \frac{1}{4\pi\eps_0}\int\limits_V \rho^\w(\vr')\frac{e^{ik|\vr-\vr'|}}{|\vr-\vr'|} dV'
\ee
\be{F32b}
\vA^\w_{i1}(\vr) = \frac{\mu_0}{4\pi}\int\limits_V \vJ_1^\w(\vr')\frac{e^{ik|\vr-\vr'|}}{|\vr-\vr'|} dV'
\ee
\end{subequations}
and
\begin{subequations}
\lbl{F33}
\be{F33a}
\vE^\w_{i2}(\vr) =  i\w \vA_{i2}^\w(\vr)
\ee
\be{F33b}
\vH^\w_{i2}(\vr) = \frac{1}{\mu_o}\nabla\times \vA_{i2}^\w(\vr)
\ee
\end{subequations}
with
\begin{subequations}
\lbl{F34}
\be{F34a}
\psi^\w_{i2}(\vr) = 0
\ee
\be{F34b}
\vA^\w_{i2}(\vr) = \frac{\mu_0}{4\pi}\int\limits_V \vJ_2^\w(\vr')\frac{e^{ik|\vr-\vr'|}}{|\vr-\vr'|} dV'
\ee
\end{subequations}
where the superscripts ``$\w$" denote the frequency-domain fields and $k=\w/c$.
\par
The volume of integration in (\ref{F22}) need cover only the PEC, which contains the charge and current, and thus the value of $|\vr-\vr'|$ in (\ref{F32}) and (\ref{F34}) does not get larger than $2a$, that is, $|\vr-\vr'|\le 2a$.  Since the PEC is assumed electrically small, $k_{\rm max}a\ll 1$, the exponentials in (\ref{F32}) and (\ref{F34}) can be expanded in a power series
\be{F35}
e^{ik|\vr-\vr'|} = 1 + ik|\vr-\vr'| - \frac{1}{2}k^2|\vr-\vr'|^2 + \ldots
\ee
and the scalar and vector potentials in (\ref{F32}) and (\ref{F34}) can be approximated by
\begin{subequations}
\lbl{F36}
\be{F36a}
\psi^\w_{i1}(\vr) = \frac{1}{4\pi\eps_0}\int\limits_V \frac{\rho^\w(\vr')}{|\vr-\vr'|} dV'  \left\{1+ O\left[(ka)^2\right]\right\}
\ee
\be{F36b}
\vA^\w_{i1}(\vr) = \frac{\mu_0}{4\pi}\int\limits_V \frac{\vJ_1^\w(\vr')}{|\vr-\vr'|} dV'\; [1+ O(ka)]
\ee
\end{subequations}
\begin{subequations}
\lbl{F37}
\be{F37a}
\psi^\w_{i2}(\vr) = 0
\ee
\be{F37b}
\vA^\w_{i2}(\vr) = \frac{\mu_0}{4\pi}\int\limits_V \frac{\vJ_2^\w(\vr')}{|\vr-\vr'|} dV' \left\{1+ O\left[(ka)^2\right]\right\}.
\ee
\end{subequations}
The order of the approximations in (\ref{F36a}) and (\ref{F37b}) are $O\left[(ka)^2\right]$ rather than $O(ka)$ because 
\begin{subequations}
\lbl{F38}
\be{F38a}
\int\limits_V \rho^\w(\vr') dV' =0
\ee
and
\be{F38b}
\int\limits_V \vJ_2^\w(\vr') dV'=0.
\ee
\end{subequations}
\par
Inserting the scalar and vector potentials from (\ref{F36}) into (\ref{F31}), the ``$i1$" frequency-domain internal electric and magnetic fields can be expressed as
\begin{subequations}
\lbl{F39}
\bea{F39a}
&&\vE^\w_{i1}(\vr) = -\frac{1}{4\pi\eps_0}\nabla\int\limits_V \frac{\rho^\w(\vr')}{|\vr-\vr'|} dV'  \left\{1+ O\left[(ka)^2\right]\right\}
+ \frac{i\w\mu_0}{4\pi}\int\limits_V \frac{\vJ_1^\w(\vr')}{|\vr-\vr'|} dV'\; [1+ O(ka)]\nonumber\\
&&= -\frac{1}{4\pi i\w\eps_0}\nabla\int\limits_V \frac{\nabla'\cdot\vJ_1^\w(\vr')}{|\vr-\vr'|} dV'  \left\{1+ O\left[(ka)^2\right]\right\}
+ \frac{i\w\mu_0}{4\pi}\int\limits_V \frac{\vJ_1^\w(\vr')}{|\vr-\vr'|} dV'\; [1+ O(ka)]\nonumber\\
&&= \frac{1}{4\pi i\w\eps_0}\left(\int\limits_V \frac{1}{R^3}\vJ_1^\w(\vr')\cdot (\vRh\vRh -\dI) dV'  \left\{1+ O\left[(ka)^2\right]\right\}
+ (ka)^2\int\limits_V \frac{\vJ_1^\w(\vr')}{a^2R} dV'\; [1+ O(ka)]\right)\nonumber\\
&&= \frac{1}{4\pi i\w\eps_0}\int\limits_V \frac{1}{R^3}\vJ_1^\w(\vr')\cdot (\vRh\vRh -\dI) dV'  \left\{1+ O\left[(ka)^2\right]\right\} = \frac{1}{4\pi\eps_0}\nabla\int\limits_V \frac{\rho^\w(\vr')}{|\vr-\vr'|} dV'  \left\{1+ O\left[(ka)^2\right]\right\}\nonumber\\
&&= \vE^\w_{es}(\vr)\left\{1+ O\left[(ka)^2\right]\right\}
\eea
\be{F39b}
\vH^\w_{i1}(\vr) = \frac{1}{\mu_o}\nabla\times \vA_{i1}^\w(\vr)[1+ O(ka)] = \frac{1}{4\pi}\nabla\times\int\limits_V \frac{\vJ_1^\w(\vr')}{|\vr-\vr'|} dV'[1+ O(ka)] = \vH^\w_{es}(\vr)[1+ O(ka)]
\ee
\end{subequations}
where $\vR = \vr-\vr'$ and the frequency-domain ``electroquasistatic fields" are
\begin{subequations}
\lbl{F40}
\be{F40a}
\vE^\w_{es}(\vr) = -\frac{1}{4\pi\eps_0}\nabla\int\limits_V \frac{\rho^\w(\vr')}{|\vr-\vr'|} dV'
\ee
\be{F40b}
\vH^\w_{es}(\vr) = \frac{1}{4\pi}\nabla\times\int\limits_V \frac{\vJ_1^\w(\vr')}{|\vr-\vr'|} dV'.
\ee
\end{subequations}
We have cavalierly ignored the ``principal volume" contribution \cite{Yaghjian-DGF}, \cite[secs. 2.3.6--2.3.7]{H&Y} to some of the volume integrals in (\ref{F39a}) because they do not change the final results in (\ref{F39}) and (\ref{F40}).
\par
Similarly, inserting the scalar and vector potentials from (\ref{F37}) into (\ref{F33}), the ``$i2$" frequency-domain internal electric and magnetic fields can be expressed as
\begin{subequations}
\lbl{F41}
\be{F41a}
\vE^\w_{i2}(\vr) =  \frac{i\w\mu_0}{4\pi}\int\limits_V \frac{\vJ_2^\w(\vr')}{|\vr-\vr'|} dV' \left\{1+ O\left[(ka)^2\right]\right\} = \vE^\w_{ms}(\vr)\left\{1+ O\left[(ka)^2\right]\right\}
\ee
\be{F41b}
\vH^\w_{i2}(\vr) = \frac{1}{\mu_o}\nabla\times \vA_{i2}^\w(\vr)\left\{1+ O\left[(ka)^2\right]\right\} = \vH^\w_{ms}(\vr)\left\{1+ O\left[(ka)^2\right]\right\} 
\ee
\end{subequations}
where the frequency-domain ``magnetoquasistatic fields" are
\begin{subequations}
\lbl{F42}
\be{F42a}
\vE^\w_{ms}(\vr) = \frac{i\w\mu_0}{4\pi}\int\limits_V \frac{\vJ_2^\w(\vr')}{|\vr-\vr'|} dV' 
\ee
\be{F42b}
\vH^\w_{ms}(\vr) = \frac{1}{4\pi}\nabla\times\int\limits_V \frac{\vJ_2^\w(\vr')}{|\vr-\vr'|} dV'.
\ee
\end{subequations}
The terms ``electroquasistatic" and ``magnetoquasistatic" were first used by Haus and Melcher \cite[ch. 3]{H&M}, although their electroquasistatic and magnetoquasistatic fields are not identical to those defined herein.  For the electrically small PEC, the magnetoquasistatic fields are produced by the magnetic dipole moment $\vm$ of the solenoidal current $\vJ_2$ (which produces no electric dipole moment), and the electroquasistatic fields are produced by the electric dipole moment $\vp$ of the charge-current $(\rho,\vJ_1)$ (which produces no magnetic dipole moment).
\par
For PEC scatterers subject to nonzero frequency-domain external fields $[\vE_e^\w,\vH_e^\w]$, the charge density $\rho^\w(\vr)$ approaches a nonzero constant $C_1(\vr) E_e^0$ as $\w \to 0$ and the irrotational current $\vJ_2^\w(\vr)$ approaches a nonzero constant vector $\vC_2(\vr)H_e^0$ as $\w \to 0$.\footnote{These conditions are merely statements that for an electrically small PEC ($k_{\rm max}a\ll1$), the charge density $\rho$ with its electric dipole moment $\vp$ is induced by the incident electric field, and the current $\vJ_2$ with its magnetic dipole moment $\vm$ is induced by the incident magnetic field.}   This allows us to obtain an order of magnitude relationship between the ``$es$" electric field in (\ref{F40a}) and the ``$ms$" electric field in (\ref{F42a}).  Specifically, as $\w\to 0$, (\ref{F40a}) and (\ref{F42a}) reveal that $\vE_{es}^\w(\vr)$ approaches a nonzero value, whereas $\vE_{ms}^\w(\vr)$ approaches zero by a factor of $\w$.  In terms of $ka$, this implies
\begin{subequations}
\lbl{F43}
\be{F43a}
\vE_{ms}^\w(\vr) = \vE_{es}^\w(\vr)O(ka)\quad {\rm as} \;\; ka\to0.
\ee
Similarly, as $\w\to0$ in a PEC scatterer, $\vJ^\w_1(\vr)$ approaches zero by a factor of $\w$ (since $\nabla\cdot\vJ^\w_1=i\w\rho^\w=i\w C_1(\vr)E_e^0$) and thus $\vH_{es}^\w(\vr)$ in (\ref{F40b}) approaches zero by a factor $\w$, whereas $\vH_{ms}^\w(\vr)$ in (\ref{F42b}) approaches a nonzero value, or in terms of $ka$
\be{F43b}
\vH_{es}^\w(\vr) = \vH_{ms}^\w(\vr)O(ka)\quad {\rm as} \;\; ka\to0.
\ee
We also have
\be{F43c}
\vJ_1^\w(\vr) = \vJ_2^\w(\vr)O(ka)\quad {\rm as} \;\; ka\to0.
\ee
\end{subequations}
\par
Combining (\ref{F39}) with (\ref{F41}) and making use of (\ref{F43a})--(\ref{F43b}) allows the near fields of electrically small PEC scatterers to be expressed in terms of the electroquasistatic and magnetoquasistatic fields in (\ref{F40}) and (\ref{F42}) as
\begin{subequations}
\lbl{FF43}
\be{FF43a}
\vE^\w_i =\vE^\w_{i1}+\vE^\w_{i2} = (\vE^\w_{es}+\vE^\w_{ms})\left\{1+ O\left[(ka)^2\right]\right\}= \vE^\w_{es}+\vE^\w_{ms},\;\;\;k_{\rm max}a\ll 1
\ee
\be{FF43b}
\vH^\w_i =\vH^\w_{i1}+\vH^\w_{i2} = \vH^\w_{es}[1+ O(ka)]+\vH^\w_{ms}\left\{1+ O\left[(ka)^2\right]\right\}= \vH^\w_{es}+\vH^\w_{ms},\;\;\;k_{\rm max}a\ll 1.
\ee
\end{subequations}
\par
Taking the Fourier transforms of (\ref{F40}) and (\ref{F42}) yields the time-domain electroquasistatic and magnetoquasistatic fields, respectively
\begin{subequations}
\lbl{F44}
\be{F44a}
\vE_{es}(\vr,t) = -\frac{1}{4\pi\eps_0}\nabla\int\limits_V \frac{\rho(\vr',t)}{|\vr-\vr'|} dV'
\ee
\be{F44b}
\vH_{es}(\vr,t) = \frac{1}{4\pi}\nabla\times\int\limits_V \frac{\vJ_1(\vr',t)}{|\vr-\vr'|} dV'
\ee
\end{subequations}
and
\begin{subequations}
\lbl{F45}
\be{F45a}
\vE_{ms}(\vr,t) = -\frac{\mu_0}{4\pi}\frac{\partial}{\partial t}\int\limits_V  \frac{\vJ_2(\vr',t)}{|\vr-\vr'|} dV' 
\ee
\be{F45b}
\vH_{ms}(\vr,t) = \frac{1}{4\pi}\nabla\times\int\limits_V \frac{\vJ_2(\vr',t)}{|\vr-\vr'|} dV'.
\ee
\end{subequations}
Similarly, from (\ref{FF43})
\begin{subequations}
\lbl{FF45}
\be{FF45a}
\vE_i(\vr,t) = \vE_{es}(\vr,t)+\vE_{ms}(\vr,t),\;\;\;k_{\rm max}a\ll 1
\ee
\be{FF45b}
\vH_i(\vr,t) = \vH_{es}(\vr,t)+\vH_{ms}(\vr,t),\;\;\;k_{\rm max}a\ll 1.
\ee
\end{subequations}
These time-domain electroquasistatic and magnetoquasistatic fields obey the differential equations
\begin{subequations}
\lbl{F46}
\be{F46a}
\nabla\times\vE_{es}(\vr,t)  = 0
\ee
\be{F46b}
\nabla\times\vH_{es}(\vr,t) -\eps_0\frac{\partial\vE_{es}(\vr,t)}{\partial t} = \vJ_1(\vr,t)
\ee
\be{F46c}
\nabla\cdot\vE_{es}(\vr,t) = \frac{\rho(\vr,t)}{\eps_0}
\ee
\be{F46d}
\nabla\cdot\vH_{es}(\vr,t) = 0
\ee
\end{subequations}
and
\begin{subequations}
\lbl{F47}
\be{F47a}
\nabla\times\vE_{ms}(\vr,t) +\mu_0\frac{\partial\vH_{ms}(\vr,t)}{\partial t} = 0
\ee
\be{F47b}
\nabla\times\vH_{ms}(\vr,t)  = \vJ_2(\vr,t)
\ee
\be{F47c}
\nabla\cdot\vE_{ms}(\vr,t) = 0
\ee
\be{F47d}
\nabla\cdot\vH_{ms}(\vr,t) = 0.
\ee
\end{subequations}
Adding these two sets of equations together gives one set of equations that, it turns out, simplifies the evaluation of the internal-field force $\vF_i(t)$ in (\ref{F22})
\begin{subequations}
\lbl{F48}
\be{F48a}
\nabla\times\vE_i(\vr,t) +\mu_0\frac{\partial\vH_{ms}(\vr,t)}{\partial t} = 0
\ee
\be{F48b}
\nabla\times\vH_i(\vr,t) -\eps_0\frac{\partial\vE_{es}(\vr,t)}{\partial t} = \vJ(\vr,t)
\ee
\be{F48c}
\nabla\cdot\vE_i(\vr,t) = \frac{\rho(\vr,t)}{\eps_0}
\ee
\be{F48d}
\nabla\cdot\vH_i(\vr,t) = 0.
\ee
\end{subequations}
We have shown that the Maxwellian equations in (\ref{F46})--(\ref{F48}){, \color{black} which predict the fields in (\ref{F44}) and (\ref{F45}) that possess no $1/r$ far fields as $r\to\infty$,} hold in the near fields of electrically small  PEC scatterers to the extent that $k_{\rm max}a \ll 1$ and, thus, as $k_{\rm max}a \to 0$, they become the exact field equations required to evaluate the internal-field force $\vF_i(t)$ in (\ref{F22}).
\subsubsection{Evaluation of the internal-field force $\vF_i(t)$}\label{EIF}
Having found the Maxwellian equations for the near fields of electrically small PEC scatterers 
in Section \ref{Nearfields}, the internal-field force in (\ref{F22}) can be evaluated starting with the electric-field part of the internal force, namely 
\be{F49}
\vF_i^E(t) =  \int\limits_{V_\infty} \rho(\vr,t)\vE_i(\vr,t) dV  = \eps_0\int\limits_{V_\infty} (\nabla\cdot\vE_i)\vE_i dV.
\ee
Use of the vector-dyadic identities
\be{F50}
(\nabla\cdot\vE_i)\vE_i =-\vE_i\cdot\nabla\vE_i +\nabla\cdot(\vE_i\vE_i) = \vE_i\times\nabla\times\vE_i -\frac{1}{2}\nabla E_i^2 +\nabla\cdot(\vE_i\vE_i)
\ee
with $E_i$ denoting the magnitude of $\vE_i$, converts (\ref{F49}) to (with the aid of (\ref{FF45a}) and (\ref{F48a}))
\be{F51}
\vF_i^E(t) = \eps_0\int\limits_{V_\infty} \vE_i\times\nabla\times\vE_i dV = -\mu_0\eps_0\int\limits_{V_\infty} \vE_i\times \frac{\partial \vH_{ms}}{\partial t}dV =-\mu_0\eps_0\int\limits_{V_\infty} (\vE_{es}+\vE_{ms})\times \frac{\partial \vH_{ms}}{\partial t}dV 
\ee
where the last two terms in (\ref{F50}) have been recast as integrals over the surface $S_\infty$ on which the fields of (\ref{F44})--(\ref{F45}) decay rapidly enough that these surface integrals are zero.  From (\ref{F46a}), $\vE_{es}=-\nabla\phi_{es}$ so that
\be{F52}
\int\limits_{V_\infty} \vE_{es}\times \frac{\partial \vH_{ms}}{\partial t}dV =-\int\limits_{V_\infty}\nabla\phi_{es} \times \frac{\partial \vH_{ms}}{\partial t}dV =\int\limits_{V_\infty}\phi_{es} \nabla\times \frac{\partial \vH_{ms}}{\partial t}dV = \int\limits_{V}\phi_{es} \times \frac{\partial \vJ_2}{\partial t}dV
\ee
after using (\ref{F47b}) and $\int_{V_\infty} \nabla\times (\phi_{es}\partial \vH_{ms}/\partial t)dV = \int_{S_\infty} \vnh\times (\phi_{es}\partial \vH_{ms}/\partial t)dS = 0$.  The $\vE_{ms}$ part of the last integral in (\ref{F51}) can be shown to equal zero as follows.
\bea{F53}
\int\limits_{V_\infty} \vE_{ms}\times \frac{\partial \vH_{ms}}{\partial t}dV =\frac{1}{\mu_0} \int\limits_{V_\infty} \vE_{ms}\times \nabla\times\frac{\partial \vA_{ms}}{\partial t}dV\hspace{65mm}\nonumber\\ = -\frac{1}{\mu_0}\int\limits_{V_\infty}\frac{\partial \vA_{ms}}{\partial t}\times \nabla\times \vE_{ms}dV - \frac{1}{\mu_0}\int\limits_{V_\infty}\frac{\partial \vA_{ms}}{\partial t}\cdot\nabla \vE_{ms}dV - \frac{1}{\mu_0}\int\limits_{V_\infty}\vE_{ms}\cdot\nabla\frac{\partial \vA_{ms}}{\partial t} dV
\eea
where use has been made of $\int_{V_\infty} \nabla (\vE_{ms}\cdot\partial \vA_{ms}/\partial t)dV = \int_{S_\infty} \vnh (\vE_{ms}\cdot\partial \vA_{ms}/\partial t)dS= 0$.  Invoking (\ref{F47a}) and (\ref{F47c}) along with the vector-dyadic identity $\nabla\cdot(\va\vb)=(\nabla\cdot\va)\vb +\va\cdot\nabla\vb$, then converting the divergence volume integrals to zero-valued surface integrals reduces (\ref{F53}) to
\be{F54}
\int\limits_{V_\infty} \vE_{ms}\times \frac{\partial \vH_{ms}}{\partial t}dV =  \int\limits_{V_\infty}\frac{\partial \vA_{ms}}{\partial t}\times\frac{\partial \vH_{ms}}{\partial t} dV 
\ee
because $\nabla\cdot\vA_{ms} =0$ as a result of $\nabla\cdot\vJ_2=0$.  The integral on the right-hand side of (\ref{F54}) is zero since 
\bea{F55}
\int\limits_{V_\infty}\frac{\partial \vA_{ms}}{\partial t}\times\frac{\partial \vH_{ms}}{\partial t} dV = \frac{1}{\mu_0}\int\limits_{V_\infty}\frac{\partial \vA_{ms}}{\partial t}\nabla\times\frac{\partial \vA_{ms}}{\partial t} dV \hspace{40mm}\nonumber\\= -\frac{1}{\mu_0}\int\limits_{V_\infty}\frac{\partial \vA_{ms}}{\partial t}\cdot\nabla\frac{\partial \vA_{ms}}{\partial t} dV = \frac{1}{\mu_0}\int\limits_{V_\infty}\frac{\partial \vA_{ms}}{\partial t}\nabla\cdot\frac{\partial \vA_{ms}}{\partial t} dV =0
\eea
in which we have utilized $\nabla\cdot\vA_{ms}=0$ and again converted the two perfect-differential volume integrals that arise to zero-valued surface integrals. 
\par
In all then, the only contribution to the electric-field part of the internal force is given from (\ref{F52}) as
\be{F56}
\vF_i^E(t) =  \int\limits_{V} \rho(\vr,t)\vE_i(\vr,t) dV  = -\mu_0\eps_0 \int\limits_{V}\phi_{es}(\vr,t) \times \frac{\partial \vJ_2(\vr,t)}{\partial t}dV.
\ee
\par
Next, we evaluate the magnetic-field part of the internal force in (\ref{F22}), namely
\be{F57}
\vF^H_{i}(t) = \mu_0\int\limits_{V_\infty} \vJ(\vr,t)\times\vH_i(\vr,t)dV = \mu_0\int\limits_{V_\infty} (\nabla\times\vH_i) \times\vH_i dV - \mu_0\eps_0\int\limits_{V_\infty}\frac{\partial \vE_{es}}{\partial t}\times\vH_i dV
\ee
in accordance with (\ref{F48b}).  The first integral on the right-hand side of (\ref{F57}) is zero; that is
\be{F58}
\int\limits_{V_\infty} (\nabla\times\vH_i) \times\vH_i dV = \int\limits_{V_\infty} \vH_i\cdot\nabla\vH_i dV = -\int\limits_{V_\infty} \vH_i\nabla\cdot\vH_i dV =0
\ee
because $\nabla\cdot\vH_i =0$.  In (\ref{F58}), the volume integrals of $\nabla H_i^2$ and $\nabla\cdot(\vH_i\vH_i)$ have been converted to zero-valued surface integrals over $S_\infty$.  This leaves
\be{F59}
\vF_i^H = - \mu_0\eps_0\int\limits_{V_\infty}\frac{\partial \vE_{es}}{\partial t}\times\vH_i dV =  \mu_0\eps_0\int\limits_{V_\infty}\nabla\frac{\partial \phi_{es}}{\partial t}\times\vH_i dV =  -\mu_0\eps_0\int\limits_{V_\infty}\frac{\partial \phi_{es}}{\partial t}\nabla\times\vH_i dV
\ee
in which the volume integral $\int_{V_\infty}\nabla\times(\vH_i \partial\phi_{es}/\partial t)dV$ has been converted to a zero-valued surface integral.  Making use of (\ref{F48b}) again recasts (\ref{F59}) into 
\be{F60}
\vF_i^H = -\mu_0\eps_0\int\limits_{V}\frac{\partial \phi_{es}}{\partial t}\vJ dV - \mu_0\eps_0^2\int\limits_{V_\infty}\frac{\partial \phi_{es}}{\partial t}\frac{\partial \vE_{es}}{\partial t} dV.
\ee
The last integral in (\ref{F60}) vanishes as a consequence of 
\be{F61}
\int\limits_{V_\infty}\frac{\partial \phi_{es}}{\partial t}\frac{\partial \vE_{es}}{\partial t} dV = -\int\limits_{V_\infty}\frac{\partial \phi_{es}}{\partial t}\nabla\frac{\partial \phi_{es}}{\partial t} dV = -\frac{1}{2}\int\limits_{V_\infty}\nabla\left(\frac{\partial \phi_{es}}{\partial t}\right)^2 dV = -\frac{1}{2}\int\limits_{S_\infty}\vnh\left(\frac{\partial \phi_{es}}{\partial t}\right)^2 dS =0
\ee
reducing the magnetic-field part of the internal force to simply
\be{F62}
\vF_i^H(t)= \mu_0\int\limits_{V} \vJ(\vr,t)\times\vH_i(\vr,t)dV = -\mu_0\eps_0\int\limits_{V}\frac{\partial \phi_{es}(\vr,t)}{\partial t}\vJ(\vr,t) dV.
\ee
\par
Adding (\ref{F62}) to (\ref{F56}) gives the total internal force as
\bea{F63}
\vF_i(t) = \vF_i^E(t)+\vF_i^H(t) = \int\limits_{V}[\rho(\vr,t)\vE_i(\vr,t) + \mu_0\vJ(\vr,t)\times\vH_i(\vr,t)]dV\nonumber\\ =  -\mu_0\eps_0 \int\limits_{V}\phi_{es}(\vr,t) \times \frac{\partial \vJ_2(\vr,t)}{\partial t}dV -\mu_0\eps_0\int\limits_{V}\frac{\partial \phi_{es}(\vr,t)}{\partial t}\vJ(\vr,t) dV.
\eea
To evaluate (\ref{F63}) for the PEC in terms of its externally applied fields, we invoke the PEC zero boundary condition on the total tangential electric field across the surface $S$ containing the current $\vJ$\footnote{If the PEC is rotating with angular velocity $\vOmega(t)$, then $\vE_{\rm tan}(\vr,t)\neq 0$ because the field $\vE(\vr,t)+\mu_0\vv(\vr,t)\times\vH(\vr,t)=0$ inside a PEC moving with velocity $\vv(\vr,t)$, which equals $\vOmega(t)\times\vr$ for the rotating PEC.  However, the rotation does not change the value of the external electric field $\vE_e(0,t)$ at the center ($\vr=0$) of the PEC.  It merely changes the magnetic dipole moment induced on the PEC by the external magnetic field.  Thus, the internal force in (\ref{F71}) remains valid for a rotating PEC as long as the angular rotation speed is restricted to $\Omega(t)\la \w_{\rm max}$ so that the rotating PEC remains electrically small with respect the wavelength of the induced fields.}
\be{F64}
\vE_{\rm tan}(\vr,t) = \left[\vE_e(\vr,t) + \vE_{es}(\vr,t) + \vE_{ms}(\vr,t)\right]_{\rm tan} =0.
\ee
However, from (\ref{F43a}), one sees that $\vE_{ms}$ is of order $ka$ times $\vE_{es}$ and thus for electrically small PEC's ($k_{\rm max}a\ll1$) the boundary condition in (\ref{F64}) becomes
\be{F65}
\left[\vE_e(\vr,t) + \vE_{es}(\vr,t) \right]_{\rm tan} = \left[\vE_e(\vr,t) - \nabla\phi_{es}(\vr,t) \right]_{\rm tan} = 0.
\ee
Expanding the external electric field in a power series about the center of the PEC's circumscribing sphere of radius $a$, as given in (\ref{F9a}), the boundary condition in (\ref{F65}) can be rewritten as
\be{F66}
\left[\vE_e(0,t) +\vr\cdot\nabla\vE_e(0,t) + \ldots - \nabla\phi_{es}(\vr,t) \right]_{\rm tan} = 0.
\ee
Integrating (\ref{F66}) between two points, $\vr_0$ and $\vr$, on the surface $S$ of the PEC yields $\phi_{es}(\vr,t)$ as
\be{F67}
\phi_{es}(\vr,t) = \vE_e(0,t)\cdot\vr + \int\limits_{\vr_0}^{\vr}\vr'\cdot\nabla\vE_e(0,t)\cdot d\vc' + \ldots ,\;\;\;\vr\in S 
\ee
where the arbitrary constant has been set equal to zero.  Under the assumption of electrically small PEC scatterers in bandlimited external fields, there is a maximum magnitude $k_{\rm max}$ of the vector spatial frequency in the spatial Fourier transform of $\vE_e(\vr,t)$ such that
\be{F68}
\left|\vr\cdot \nabla\vE_e(0,t)\right|_{\rm max} = |\vE_e(0,t)|O(k_{\rm max}a).
\ee
In addition, the line integration in (\ref{F67}) covers a distance on the order of $a$.  Thus, (\ref{F67}) can be expressed in the form
\be{F69}
\phi_{es}(\vr,t) = \vE_e(0,t)\cdot\vr + |\vE_e(0,t)|a O(k_{\rm max}a),\;\;\;\vr\in S. 
\ee
Since the first term on the right-hand side of (\ref{F69}) is of order $|\vE_e(0,t)|a$, the second term on the right-hand side of (\ref{F69}) is negligible for electrically small scatterers ($k_{\rm max}a\ll 1$) and (\ref{F63}) can be re-expressed as
\be{F70}
\vF_i(t) = \vF^E_i(t)+\vF^H_i(t) = -\mu_0\eps_0 \vE_e(0,t)\cdot\frac{d}{dt}\int\limits_{V}\vr\vJ_2(\vr,t) dV -\mu_0\eps_0\frac{\partial\vE_e(0,t)}{\partial t}\cdot\int\limits_{V}\vr\vJ(\vr,t) dV.
\ee
\par
The integral expressions in (\ref{F70}) can be recast in terms of the magnetic dipole moment of the PEC by using the vector-dyadic identities in (\ref{F16b})--(\ref{F16c}) and the definition of the magnetic dipole moment in (\ref{F18'}) and (\ref{F26-m}) to get
\be{F71} 
\vF_i(t) =  \vF^E_i(t)+\vF^H_i(t) = -\mu_0\eps_0 \frac{d\vm(t)}{dt}\times\vE_e(0,t) -\mu_0\eps_0\vm(t)\times\frac{\partial\vE_e(0,t)}{\partial t} = -\frac{1}{c^2}\frac{\partial}{\partial t}\left[\vm(t)\times\vE_e(0,t)\right]
\ee
where we have also made use of $\int_V(\vr\vJ_2+\vJ_2\vr)dV/2 = (d/dt)\int_V \nabla\cdot\vJ_2 \vr\vr dV = 0$ because $\nabla\cdot\vJ_2=0$ and $\int_V(\vr\vJ+\vJ\vr)dV/2 = (d/dt)\int_V \nabla\cdot\vJ \vr\vr dV = d\dQ/dt = 0$ because the quadrupole moment of an electrically small enough PEC is negligible compared to the electric and magnetic dipole moments.  The equation in (\ref{F71}) expresses the significant result that a time-varying microscopic Amperian magnetic dipole, represented classically by an electrically small PEC in time-varying external fields, exhibits (by means of a rigorous solution to Maxwell's equations for the PEC boundary-value problem) an internal force exerted on the charge-current by the internal fields induced in the PEC by the external fields.  This internal force exactly equals the time rate of change of the proverbial hidden momentum originally introduced by Shockley and James \cite{S&J}.  The derivation of the internal force in (\ref{F71}) requires no restrictive assumptions other than the PEC be electrically small {\color{black} and the externally applied fields be bandlimited} ($k_{\rm max}a\ll1$).  {\color{black}Otherwise}, the fields and charge-currents can have arbitrary time variation as long as the PEC remains electrically small.  Both the externally applied fields and the induced fields on the PEC can include radiation fields (and thus they need not satisfy the requirements of the conservation theorems that Coleman and Van Vleck \cite{C&V} use to substantiate the claims of Shockley and James). 
\subsection{Total force on the charge-current of the PEC}\label{TF}
The internal force in (\ref{F71}) added to the external force in (\ref{F21}) gives the total force on the charge-current of the %stationary 
electrically small PEC centered at $\vr=0$ in external fields $[\vE_e(\vr,t),\vH_e(\vr,t)]$ as
\be{F72}
\vF(t) = \vF_e(t)+\vF_i(t)=\vp(t)\cdot\nabla\vE_e(0,t) + \mu_0 \frac{d\vp(t)}{dt}\times\vH_e(0,t) + \mu_0\vm(t)\cdot\nabla\vH_e(0,t) - \mu_0\eps_0\frac{d\vm(t)}{dt}\times\vE_e(0,t)
\ee
which is identical to the force on a microscopic electric-charge electric dipole moment $\vp(t)$ and a magnetic-charge magnetic dipole moment $\vm(t)$ in the same externally applied fields \cite[eq. (2.166)]{H&Y}.\footnote{The force in (\ref{F72}) for magnetic-charge magnetic dipoles (and electric-charge electric dipoles) is obtained in \cite[eq. (2.166)]{H&Y} by applying the same procedure to evaluate $\int_V(\rho\vE_e +\mu_0\vJ\times\vH_e + \rho_m\vE_e +\mu_0\vJ_m\times\vH_e)dV$ that was used to evaluate (\ref{F8}) in Section \ref{FEEF} with the electric dipole moment produced by the electric charge-current ($\rho,\vJ$) but now with the magnetic dipole moment produced by the magnetic charge-current ($\rho_m,\vJ_m$).  Assuming a model of an electric dipole as oppositely moving positive and negative charges, such as equal and opposite charges sliding on a rod that can be rotating about a fixed center, the internal forces between the electric charges cancel \cite[eqs. (2.99)--(2.100)]{H&Y}.  Likewise, the internal forces cancel between the equal and opposite magnetic charges of a rotating-rod model of a magnetic dipole.  There may be a force on the isolated microscopic electric-charge electric dipole from the fields of the isolated microscopic magnetic-charge magnetic dipole and vice versa, but these two forces can be taken into account merely by changing the value of the effective external fields in the electric and magnetic dipole-moment terms, respectively, of (\ref{F72}).  (If one divides the derivation into two parts by separately finding the forces on electric-charge electric dipoles alone and on magnetic-charge magnetic dipoles alone, this issue of the forces between isolated microscopic electric and magnetic dipoles does not arise because the external fields in each of the separate derivations can be chosen independently.  Interestingly, the electric and magnetic dipole moments of a PEC are not isolated and yet the total force they experience is given by (\ref{F72}).)} Thus, we have proven, using classical PEC scatterers, that the force on microscopic electric-charge-current electric and magnetic dipoles in external fields contains an internal (``hidden-momentum") force that makes the total force identical to that on microscopic electric-charge electric dipoles and magnetic-charge magnetic dipoles.
\par
If we assume that the PEC is composed of electrons moving freely in a fixed rigid lattice of positive charge, the force in (\ref{F72}) on the charge-current of %a stationary (fixed) 
the PEC can manifest itself in two ways. First, the force can be transferred to the fixed lattice of the PEC which, in turn, can be transferred to the structure holding the lattice fixed. Second, it can change the kinetic momentum of the electrons.  However, in a classical model of a PEC that instantaneously maintains zero fields within the conductor, the mass of the charge carriers (electrons) has to be negligible and thus the energy and momentum of the charge carriers is negligible except possibly for the charge carriers moving at velocities approaching the speed of light.  It is this latter possibility that occurs in the Penfield-Haus model \cite{P&H}, discussed above in Section \ref{Cm} of the Introduction, of an Amperian magnetic dipole consisting of charges circulating in a frictionless tube subject to an applied external electric field.  In the Penfield-Haus model, even as the mass of the charge carriers approaches zero, the velocity of the charge carriers approaches the speed of light such that the applied electric field induces a change of relativistic kinetic energy that leads to a net relativistic kinetic momentum as the charge carriers circulate in the tube --- this relativistic kinetic momentum being equal to the ``hidden momentum" of the Amperian magnetic dipole.  Emphatically, this is not the case for a PEC where the tangential electric field across the surface current of the PEC remains zero in an externally applied field and, thus, the charge carriers experience no change in their kinetic energy or kinetic momentum as they circulate in the external field.  In other words, the zero tangential boundary condition across the surface current of a stationary PEC eliminates the possibility that part of the hidden momentum of the Amperian magnetic dipole is exhibited as kinetic momentum of the charge carriers.  All of the hidden momentum of a rigid PEC is exhibited as a force on the structure to which the PEC is fixed.  In this respect, the PEC is similar to Boyer's model \cite{Boyer}, \cite{BoyerAJP}, discussed above in Section \ref{Cm} of the Introduction, consisting of two low-velocity interacting charges circulating around a circular ring path with an opposite compensating charge at the center of the ring.
\section{FORCE ON A PEC SPHERE ILLUMINATED BY A PLANE WAVE}\label{PECSphere}
In order to verify the expression in (\ref{F72}) for the total electromagnetic force on an electrically small PEC, we determine the force on a rigid PEC sphere in a plane-wave external field from the exact Mie solution as the radius $a$ of the exterior surface of the sphere becomes electrically small ($ka\ll1$).  Given the relatively simple, straightforward expressions for the fields in the Mie solution \cite[sec. 9.25]{Stratton}, it may seem somewhat surprising that this determination of the time-domain force on a PEC sphere has not been determined previously (as far as I am aware).  One reason for this is that to determine the time-domain force on the sphere in the plane-wave field, the real fields must be considered and not just the complex phasor fields.  Another reason may be that usually only the dipole fields are kept in the Mie solution as $ka\to 0$ since the multipole moments and their far fields of higher order than dipoles are negligible compared with the electric and magnetic dipole moments and their far fields as $ka\to 0$.  However, it turns out, as shown below, that even though the ratios of the moments and far fields of the higher-order multipoles to those of the dipoles approach zero as $ka\to 0$, the near fields at the surface of the sphere of the electric and magnetic quadrupoles must be retained in the Mie solution to obtain the correct time-domain force on the dipoles of the PEC sphere. 
\par
The total electromagnetic force on the PEC sphere is given by the expression in {\color{black}(\ref{S2})} or (\ref{F7}), namely
\be{F73} 
\vF(t) = \int\limits_V [\rho(\vr,t)\vE(\vr,t) + \mu_0\vJ(\vr,t)\times\vH(\vr,t)]dV 
\ee
with the illuminating external plane-wave fields given by
\begin{subequations}
\lbl{F74}
\be{F74a}
\vE_e(\vr,t) =\vxh{\rm Re}[E^0 e^{i(\w t-kz)}] = \vxh E^0 \cos(\w t-kz)
\ee
\be{F74b}
\vH_e(\vr,t) =\vyh{\rm Re}[H^0 e^{i(\w t-kz)}] = \vyh H^0 \cos(\w t-kz).
\ee
\end{subequations}
The superscripts ``0" (instead of subscripts) are used here to distinguish the constants ($E^0,H^0$) with $E^0 = Z_0H^0$ and $Z_0=\sqrt{\mu_0/\eps_0}$  from the symbols used previously for the static fields.  The rectangular coordinates ($x,y,z$) have associated spherical coordinates ($r,\theta,\phi$) so that $z=r\cos\theta$ in (\ref{F74}).
\subsection{Force exerted by the total electric field on the charge density of the PEC sphere}\label{FE}
The force $\vF^E(t)$ exerted by the total electric field $\vE(\vr,t)$ on the induced charge density $\rho(\vr,t)$ of the sphere can be determined from 
\be{F75} 
\vF^E(t) = \int\limits_V \rho(\vr,t)\vE(\vr,t)dV = \int\limits_S \vrh\cdot\dT^E(\vr,t)dS -\frac{1}{c^2}\int\limits_V \vE(\vr,t)\times\frac{\partial\vH(\vr,t)}{\partial t} dV
\ee
where the electric stress dyadic is
\be{F76}
\dT^E(\vr,t) =\eps_0\left[\vE(\vr,t)\vE(\vr,t)-\frac{1}{2}\dI E^2(\vr,t)\right].
\ee
Let $V$ be the spherical volume of radius $a$ such that the surface $S$ of $V$ just encloses all the charge-current on the PEC sphere.  Then the volume integral on the right-hand side of (\ref{F75}) vanishes because the total fields are zero inside the PEC sphere and finite throughout the infinitesimally thin surface layer of charge-current.  Thus, (\ref{F75}) reduces to
\be{F77} 
\vF^E(t) = \int\limits_{V_a} \rho(\vr,t)\vE(\vr,t)dV = \int\limits_S \vrh\cdot\dT^{E}(\vr,t) dS
\ee
where $\dT^E(\vr,t)$ is evaluated on $S$ just outside the PEC sphere.  Because  $\vE_{t} =0$ on $S$, (\ref{F77}) combines with (\ref{F76}) to give
\be{F78} 
{\color{black}\vF^E(t) = \frac{\eps_0}{2} \int\limits_S E_{r}^2\vrh dS =\frac{\eps_0 a^2}{2}\int\limits_0^{2\pi} \int\limits_0^{\pi}E_{r}^2\vrh\sin\theta d\theta d\phi =\frac{\eps_0 a^2}{2}\int\limits_0^{2\pi} \int\limits_0^{\pi}(E_{e r} + E_{s r})^2\vrh\sin\theta d\theta d\phi}
\ee
with the subscripts ``$r$" and ``$t$" denoting the radial (normal) and tangential vector components with respect to the spherical surface $S$.  As in previous sections, the subscript ``$e$" refers to the external fields and the subscript ``$s$" refers to the scattered fields produced by the charge and current induced on the sphere by the external fields.
\par
To evaluate the double integral in (\ref{F78}), we can insert the fields from the Mie solution in \cite[p. 564]{Stratton} with the coefficients in \cite[eq.(13) on p. 565]{Stratton} for the PEC sphere.  However, only the terms of order $a^3$ need be retained in (\ref{F78}) because the electric and magnetic dipole moments, which are the lowest order multipole moments, are of order $a^3$ and thus all higher order terms become negligible as the sphere radius approaches zero, in particular, for electrically small spheres.  In other words, the force on the multipole moments of higher order than dipole moments approaches zero as $ka$ approaches zero.  This means that only the portion of the fields to order $ka$ need be retained under the integral signs in (\ref{F78}) for electrically small spheres.  Specifically, we have from (\ref{F74a})
\be{F79}
E_{e r} = E^0 \sin\theta\cos\phi \cos(ka\cos\theta - \w t) \approx E^0\sin\theta\cos\phi(\cos\w t + ka\cos\theta\sin\w t)
\ee
and from \cite[p. 564]{Stratton}
\be{F80}
E_{sr} \approx E^0 \vrh\cdot{\rm Re}\left[\left(\frac{3}{2}b_1\vn_{{\rm e}1}+\frac{5}{6}ib_2\vn_{{\rm e}2} \right)e^{-i\w t}\right]
\approx E^0\sin\theta\cos\phi\left(2\cos \w t +\frac{3}{2}ka\cos\theta\sin\w t \right)
\ee
where $\vn_{{\rm e}1}$ and $\vn_{{\rm e}2}$ are the Stratton ``even" electric dipole and electric quadrupole exterior electric fields, respectively, and $b_1$ and $b_2$ are their coefficients, which are functions of $ka$. (Subscripts ``$1$" denoting $m=1$ that are common to all the $\vm$ and $\vn$ functions of the Stratton Mie solution are omitted, as well as the superscripts ``$r$" denoting ``reflected" on the Stratton $b_n$'s.)  Embedded in $\vn_{{\rm e}1}$ and $\vn_{{\rm e}2}$ are the spherical Hankel functions $h_1^{(1)}(ka)$ and $h_2^{(1)}(ka)$, respectively.  To obtain the last approximate expression in (\ref{F80}), use has been made of the small $ka$ approximations
\begin{subequations}
\lbl{F81}
\be{F81a}
\frac{b_1 h_1^{(1)}(ka)}{ka} \stackrel{ka\to0}{\sim} \frac{2}{3} +O[(ka)^2]
\ee
\be{F81b}
{\color{black}\frac{b_2 h_2^{(1)}(ka)}{ka} \stackrel{ka\to0}{\sim} \frac{1}{10}ka +O[(ka)^3]}
\ee
\end{subequations}
found from Stratton's Mie solution for the PEC sphere.
{\color{black}In view of (\ref{F79}) and (\ref{F80}), $E_r=E_{er}+E_{sr} \approx E^0\sin\theta\cos\phi [3\cos\w t +(5/2)ka\cos\theta\sin\w t]$ and
\be{F82}
E_r^2 = (E_{e r} + E_{s r})^2 \approx (E^0)^2 \sin^2\theta\cos^2\phi\,(9\cos\w t + 15ka\cos\theta\sin\w t)\cos \w t.
\ee
Inserting (\ref{F82}) under the last double integral sign of (\ref{F78}) and performing the integrations shows that only the $ka$ term in (\ref{F82}) multiplied by the $z$ component of $\vrh$ survives to give}
\be{F83}
\vF^E(t) =  2\pi\eps_0 ka^3 (E^0)^2 \sin\w t \cos \w t \,\vzh\,.
\ee
For electrically small spheres ($ka\to 0$), the magnetic dipole moment of the PEC sphere approaches the value of
\be{F84}
\vm(t) = -2\pi a^3 H^0\cos \w t\,\vyh
\ee
so that
\be{F85}
\frac{d\vm(t)}{dt} = 2\pi\w a^3 H^0\sin\w t\,\vyh.
\ee
The combination of (\ref{F85}) and (\ref{F74a}) reveals that $\vF^E(t)$ in (\ref{F83}) can be rewritten as
\be{F86}
\vF^E(t) =  -\mu_0\eps_0\frac{d\vm(t)}{dt}\times\vE_e(0,t).
\ee
This force can also be written as the sum of the external electric-field force and the internal electric-field force
\be{F87}
\vF^E(t)=\vF_e^E(t)+\vF_i^E(t)
\ee
and, thus, we see that (\ref{F86}) agrees with the general results in (\ref{F12}) plus the $\vF_i^E$ part of (\ref{F71}) for the electric-field force on electrically small PEC scatterers because $\vp(t)\cdot\nabla\vE_e(0,t)=0$ for the PEC sphere under plane-wave illumination.  It is noteworthy that the force in (\ref{F86}) requires the electric quadrupolar electric field in (\ref{F80}) at the surface of the sphere, even though the electric quadrupole moment and far fields are negligible compared to the electric and magnetic dipole moments and far fields for electrically small spheres (that is, their ratios approach zero as $ka\to0$).
\par
If we assume that the PEC sphere maintains a static charge density (produced, for example, by a uniform static external electric field) that gives rise to a static electric dipole moment $\vp_0$, then the modification of the above analysis that includes these static electric dipole fields adds a force to (\ref{F86}) equal to $\vp_0\cdot\nabla\vE_e(0,t)$, that is
\be{F88}
\vF^E(t) = \vp_0\cdot\nabla\vE_e(0,t)  -\mu_0\eps_0\frac{d\vm(t)}{dt}\times\vE_e(0,t)
\ee
which again agrees with the general results in (\ref{F12}) plus the $\vF_i^E$ part of (\ref{F71}) for the electric-field force on electrically small PEC scatterers.
\subsection{Force exerted by the total magnetic field on the current density of the PEC sphere}\label{FH}
The force $\vF^H(t)$ exerted by the total  field $\vH(\vr,t)$ on the induced current density $\vJ(\vr,t)$ of the sphere is determined from
\be{F89} 
\vF^H(t) = \mu_0 \int\limits_V \vJ(\vr,t)\times\vH(\vr,t)dV = \int\limits_S \vrh\cdot\dT^H(\vr,t)dS -\frac{1}{c^2}\int\limits_V \frac{\partial\vE(\vr,t)}{\partial t}\times\vH(\vr,t) dV
\ee
where the magnetic stress dyadic is
\be{F90}
\dT^H(\vr,t) =\mu_0\left[\vH(\vr,t)\vH(\vr,t)-\frac{1}{2}\dI H^2(\vr,t)\right].
\ee
As we did for $\vF^E(t)$ in the previous subsection, $\vF^H(t)$ in (\ref{F89}) can be reduced to
\be{F91} 
\vF^H(t) = \mu_0\int\limits_{V_a} \vJ(\vr,t)\times\vH(\vr,t)dV = \int\limits_S \vrh\cdot\dT^{H}(\vr,t) dS.
\ee
Because $H_{r} =0$ on $S$, (\ref{F91}) combines with (\ref{F90}) to give
\be{F92} 
{\color{black}\vF^H(t) = -\frac{\mu_0}{2} \int\limits_S |\vH_{t}|^2\vrh dS = -\frac{\mu_0 a^2}{2}\int\limits_0^{2\pi} \int\limits_0^{\pi}|\vH_{t}|^2\vrh\sin\theta d\theta d\phi = -\frac{\mu_0 a^2}{2}\int\limits_0^{2\pi} \int\limits_0^{\pi}|\vH_{e t} + \vH_{s t}|^2\vrh\sin\theta d\theta d\phi.}
\ee
\par
To evaluate the double integral in (\ref{F92}), we can insert the fields from the PEC Mie solution in \cite[pp. 564--565]{Stratton}.  However, as explained above for $\vF^E(t)$, only the portion of the fields to order $ka$ need be retained under the integral signs in (\ref{F92}) for electrically small spheres.  Specifically, we have from (\ref{F74b})
\be{F93}
\vH_{et} \approx H^0 (\cos\theta\sin\phi\,\vthh + \cos\phi\,\vphih)(\cos\w t + ka\cos\theta\sin\w t)
\ee
and from \cite[p. 564]{Stratton}
\bea{F94}
\vH_{st} \approx -H^0 {\rm Re}\left\{\left[\frac{3}{2}(ib_1\vm_{{\rm e}1} - a_1\vn_{{\rm o}1}) -\frac{5}{6}ia_2\vn_{{\rm o}2}\right]_t e^{-i\w t}\right\}\nonumber\hspace{60mm}\\ \approx H^0\left\{\frac{1}{2}(\cos\theta\sin\phi\,\vthh + \cos\phi\,\vphih)\cos\w t +\frac{2}{3}ka\left[\sin\phi(1+\cos^2\theta)\,\vthh +2\cos\theta\cos\phi\,\vphih\right]\sin\w t\right\}
\eea
where $\vm_{{\rm e}1}$ is the Stratton``even" electric dipole exterior magnetic field.  The $\vn_{{\rm o}1}$ and $\vn_{{\rm o}2}$ are the Stratton ``odd" magnetic dipole and magnetic quadrupole exterior magnetic fields, respectively, and $b_1$, $a_1$ and $a_2$ are their coefficients, which are functions of $ka$. (Subscripts ``$1$" denoting $m=1$ that are common to all the $\vm$ and $\vn$ functions of the Stratton Mie solution are omitted, as well as the superscripts ``$r$" denoting ``reflected" on the Stratton $a_n$'s and $b_n$'s.)  Embedded in [$\vm_{{\rm e}1}$, $\vn_{{\rm o}1}$] and $\vn_{{\rm o}2}$ are the spherical Hankel functions $h_1^{(1)}(ka)$ and $h_2^{(1)}(ka)$, respectively.  To obtain the last approximate expression in (\ref{F94}), use has been made of the small $ka$ approximation in (\ref{F81a}) as well as the small $ka$ approximations
\begin{subequations}
\lbl{F95}
\be{F95a}
a_1\frac{1}{ka} \frac{d}{d(ka)}\left[ka h_1^{(1)}(ka)\right] \stackrel{ka\to0}{\sim} \frac{1}{3} +O[(ka)^2]
\ee
\be{F95b}
{\color{black}a_2 \frac{1}{ka}\frac{d}{d(ka)}\left[ka h_2^{(1)}(ka)\right] \stackrel{ka\to0}{\sim} \frac{2}{15}ka +O[(ka)^3]}
\ee
\end{subequations}
found from Stratton's Mie solution.
In view of (\ref{F93}) and (\ref{F94})
\be{F96}
|\vH_{e t} + \vH_{s t}|^2 \approx (H^0)^2\left[\frac{9}{4}(1-\sin^2\theta\sin^2\phi)\cos^2\w t +ka\cos\theta(7-5\sin^2\theta\sin^2\phi)\sin\w t \cos\w t\right].
\ee
{\color{black}Inserting (\ref{F96}) under the last double integral sign of (\ref{F92}) and performing the integrations shows that only the $ka$ term in (\ref{F96}) multiplied by the $z$ component of $\vrh$ survives to give}
\be{F97}
\vF^H(t) =  -4\pi\mu_0 ka^3 (H^0)^2 \sin\w t \cos \w t \,\vzh\,.
\ee
For electrically small spheres ($ka\to 0$), the electric dipole moment of the PEC sphere approaches the value of
\be{F98}
\vp(t) = 4\pi\eps_0 a^3 E^0\cos \w t\,\vxh
\ee
so that
\be{F99}
\frac{d\vp(t)}{dt} = -4\pi\w a^3 E^0\sin\w t\,\vxh.
\ee
The combination of (\ref{F99}) and (\ref{F74b}) reveals that $\vF^H(t)$ in (\ref{F97}) can be rewritten as
\be{F100}
\vF^H(t) =  \mu_0\frac{d\vp(t)}{dt}\times\vH_e(0,t).
\ee
This force can also be written as the sum of the external magnetic-field force and the internal magnetic-field force
\be{F101}
\vF^H(t)=\vF_e^H(t)+\vF_i^H(t)
\ee
and, thus, we see that (\ref{F100}) agrees with the general results in (\ref{F20'}) plus the $\vF_i^H$ part of (\ref{F71}) for the magnetic-field force on electrically small PEC scatterers because $\vm(t)\cdot\nabla\vH_e(0,t)=0$ for the PEC sphere under plane-wave illumination.  It is noteworthy that the force in (\ref{F100}) requires the magnetic quadrupolar magnetic field in (\ref{F94}) at the surface of the sphere, even though the magnetic quadrupole moment and far fields are negligible compared to the electric and magnetic dipole moments and far fields for electrically small spheres (that is, their ratios approach zero as $ka\to0$).
\par
If we assume that the PEC sphere maintains a static current density (produced, for example, by a uniform static external magnetic field) that gives rise to a static magnetic dipole moment $\vm_0$, then the modification of the above analysis that includes these static magnetic dipole fields adds a force to (\ref{F100}) equal to $\mu_0\vm_0\cdot\nabla\vH_e(0,t)$, that is
\be{F102}
\vF^H(t) = \mu_0\vm_0\cdot\nabla\vH_e(0,t)  +\mu_0\frac{d\vp(t)}{dt}\times\vH_e(0,t)
\ee
which again agrees with the general results in (\ref{F20'}) plus the $\vF_i^H$ part of (\ref{F71}) for the magnetic-field force on electrically small PEC scatterers.
\par
Adding $\vF^E(t)$ in (\ref{F88}) to $\vF^H(t)$ in (\ref{F102}) gives the total electromagnetic force exerted on the charge-current of the electrically small PEC sphere by the external plane-wave illumination in (\ref{F74})
\be{F103}
\vF(t) = \vp_0(t)\cdot\nabla\vE_e(0,t) + \mu_0 \frac{d\vp(t)}{dt}\times\vH_e(0,t) + \mu_0\vm_0(t)\cdot\nabla\vH_e(0,t) - \mu_0\eps_0\frac{d\vm(t)}{dt}\times\vE_e(0,t)
\ee
which checks with the total force in (\ref{F72}) on the charge-current of an electrically small PEC of arbitrary shape centered at $\vr=0$ in an arbitrary bandlimited time-varying external fields [$\vE_e(\vr,t),\vH_e(\vr,t)$].  The agreement between the two expressions in (\ref{F103}) and (\ref{F72}) mutually confirms the analyses used to derive each of them and further substantiates that a microscopic Amperian magnetic dipole contains an internal (``hidden-momentum") force that makes the total force exerted on their charge-current {\color{black} in} the external fields identical to the force on a microscopic magnetic-charge magnetic dipole with the same magnetic dipole moment in the same external fields.
\section{CONCLUSION}\lbl{Conclusion}
After a selective review of ``hidden momentum" and the various approximate methods and arguments that have been used for determining the electromagnetic force on Amperian magnetic dipoles, we rigorously solve Maxwell's equations for the force on electrically small perfect electric conductors (PEC's) carrying time-varying electric and magnetic dipoles induced by time-varying external fields.  We prove unequivocally that there is an internal (``hidden-momentum") force exerted on the electric charge-current of the PEC by internal fields in the PEC produced by the electric charge-current --- thereby making the force on the Amperian magnetic dipole equal to the force on a magnetic-charge magnetic dipole with the same magnetic dipole moment in the same external field.  Furthermore, it is shown that these electromagnetic forces exerted on the charge-current of the rigid PEC are transferred to the structure holding the PEC fixed and none of it is manifested as a change in kinetic momentum of the charge carriers.
\par
The force on the dipoles exerted directly by the externally applied fields is derived straightforwardly with the help of power series expansions for the external fields.  The derivation of the more elusive internal force on the dipoles exerted by the fields of the electric charge-current on itself is facilitated by first dividing the electric currents into solenoidal and nonsolenoidal parts, with each part compactly supported by the PEC, then finding the Maxwellian equations satisfied by the fields of each of these currents that flow on the electrically small PEC.
\par
The expressions obtained for the electromagnetic force on the dipoles of electrically small PEC's are confirmed by finding the exact time-domain electromagnetic force on an electrically small rigid PEC sphere illuminated by a plane wave (the Mie solution).  Rather remarkably, the derivation of the force on the PEC sphere reveals that the electric and magnetic quadrupolar fields at the surface of the sphere are required to obtain the correct hidden-momentum force on the Amperian magnetic dipoles, even though the electric and magnetic quadrupole moments and their far fields are negligible compared to the electric and magnetic dipole moments and their far fields (that is, their ratios approach zero as the electrical size of the PEC sphere approaches zero).
%\par
%Lastly, a general argument based on Newton's third law of motion applied to the mutual force between sources lying in each others quasistatic fields (and involving relatively little mathematical analysis) is used to show that the hidden momentum, which makes the force exerted on Amperian magnetic dipoles equal to the force on magnetic-charge magnetic dipoles in the same externally applied fields, exists not only for the Amperian magnetic dipoles of PEC's but also for all other passive, lossy or lossless, microscopic Amperian magnetic dipoles.  The Mie solution for electrically small lossy spheres confirms that the same expression for hidden momentum applies to both lossy and lossless passive microscopic Amperian magnetic dipoles.
% 
%\section*{{Acknowledgements}}
%\begin{acknowledgments}
\ack
This research was supported under the U.S. Air Force Office of Scientific Research (AFOSR) Grant \# FA9550-16-C-0017 through Dr. A. Nachman. %Helpful discussions with Professors M.G. Silveirinha and A. Al\`u are gratefully acknowledged.
%
%\end{acknowledgments}
%
%
%

%
%
\end{document}